\documentclass[nojss]{jss}
\shortcites{Hamill+Bates+Whitaker:2013}

\usepackage{amstext,amsfonts,amsmath,bm,thumbpdf,lmodern,hyperref}
\usepackage[all]{hypcap}

\usepackage{array,makecell,tikz,color,soul}
\usetikzlibrary{arrows.meta,positioning,shapes,arrows,decorations.pathreplacing,calc,automata,mindmap}

\newcommand{\argmin}{\operatorname{argmin}\displaylimits}
\newcommand{\argmax}{\operatorname{argmax}\displaylimits}

\newcommand{\I}{\mathbf{I}}

\definecolor{changes}{rgb}{1,1,0.7}

\numberwithin{equation}{subsection}

\title{Distributional Regression Forests for Probabilistic Precipitation Forecasting in Complex Terrain}
\Shorttitle{Distributional Regression Forests for Probabilistic Precipitation Forecasting}

\author{Lisa Schlosser\\Universit\"at Innsbruck
   \And Torsten Hothorn\\Universit\"at Z\"urich
   \And Reto Stauffer\\Universit\"at Innsbruck
   \And Achim Zeileis\\Universit\"at Innsbruck}
\Plainauthor{Lisa Schlosser, Torsten Hothorn, Reto Stauffer, Achim Zeileis}

\Abstract{
To obtain a probabilistic model for a dependent variable based on some set of
explanatory variables, a distributional approach is often adopted where the
parameters of the distribution are linked to regressors. In many classical
models this only captures the location of the distribution but
over the last decade there has been increasing interest in distributional
regression approaches modeling all parameters including location, scale, and
shape. Notably, so-called non-homogeneous Gaussian regression (NGR) models both
mean and variance of a Gaussian response and is particularly
popular in weather forecasting. 
Moreover, generalized additive models for location, scale, and shape (GAMLSS)
provide a framework where each distribution parameter is modeled separately
capturing smooth linear or nonlinear effects.
However, when variable selection is required and/or there are
non-smooth dependencies or interactions (especially unknown or of high-order),
it is challenging to establish a good GAMLSS. A natural alternative in these
situations would be the application of regression trees or random forests but,
so far, no general distributional framework is available for these. Therefore,
a framework for distributional regression trees and forests is proposed that blends
regression trees and random forests with classical distributions from the GAMLSS
framework as well as their censored or truncated counterparts.
To illustrate these novel approaches in practice, they are employed to obtain
probabilistic precipitation forecasts at numerous sites in a mountainous
region (Tyrol, Austria) based on a large number of numerical weather
prediction quantities. It is shown that the novel distributional regression forests
automatically select variables and interactions, performing on par or often even
better than GAMLSS specified either through prior meteorological knowledge or a
computationally more demanding boosting approach.
}

\Address{
  Lisa Schlosser, Reto Stauffer, Achim Zeileis \\
  Universit\"at Innsbruck \\
  Department of Statistics \\
  Faculty of Economics and Statistics \\
  Universit\"atsstr.~15 \\
  6020 Innsbruck, Austria \\
  E-mail: \email{Lisa.Schlosser@uibk.ac.at}, \email{Reto.Stauffer@uibk.ac.at},\\
  \email{Achim.Zeileis@R-project.org} \\
  URL: \url{https://www.uibk.ac.at/statistics/personal/schlosser-lisa/},\\
  \phantom{URL: }\url{https://retostauffer.org/},\\
  \phantom{URL: }\url{https://eeecon.uibk.ac.at/~zeileis/}\\
  
  Torsten Hothorn\\
  Universit\"at Z\"urich \\
  Institut f\"ur Epidemiologie, Biostatistik und Pr\"avention \\
  Hirschengraben 84\\
  CH-8001 Z\"urich, Switzerland \\
  E-mail: \email{Torsten.Hothorn@R-project.org}\\
  URL: \url{http://user.math.uzh.ch/hothorn/}\\
  
}

\begin{document}

\section{Introduction}
\label{sec:introduction}

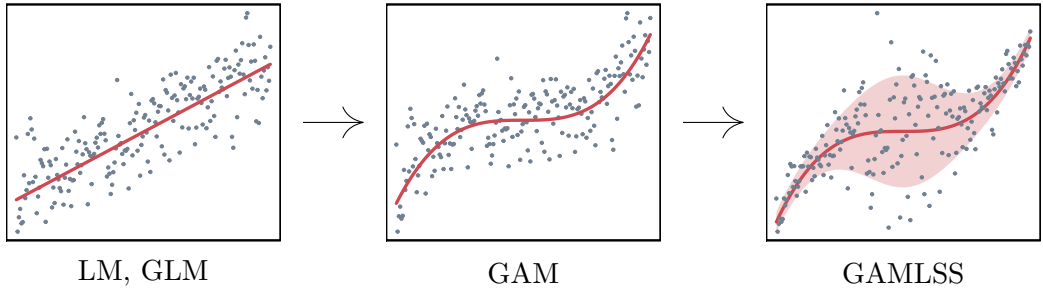
\begin{figure}[t!]
\begin{tikzpicture}
\node (a) at (0,0){
\minipage{0.29\textwidth}
\centering

\includegraphics{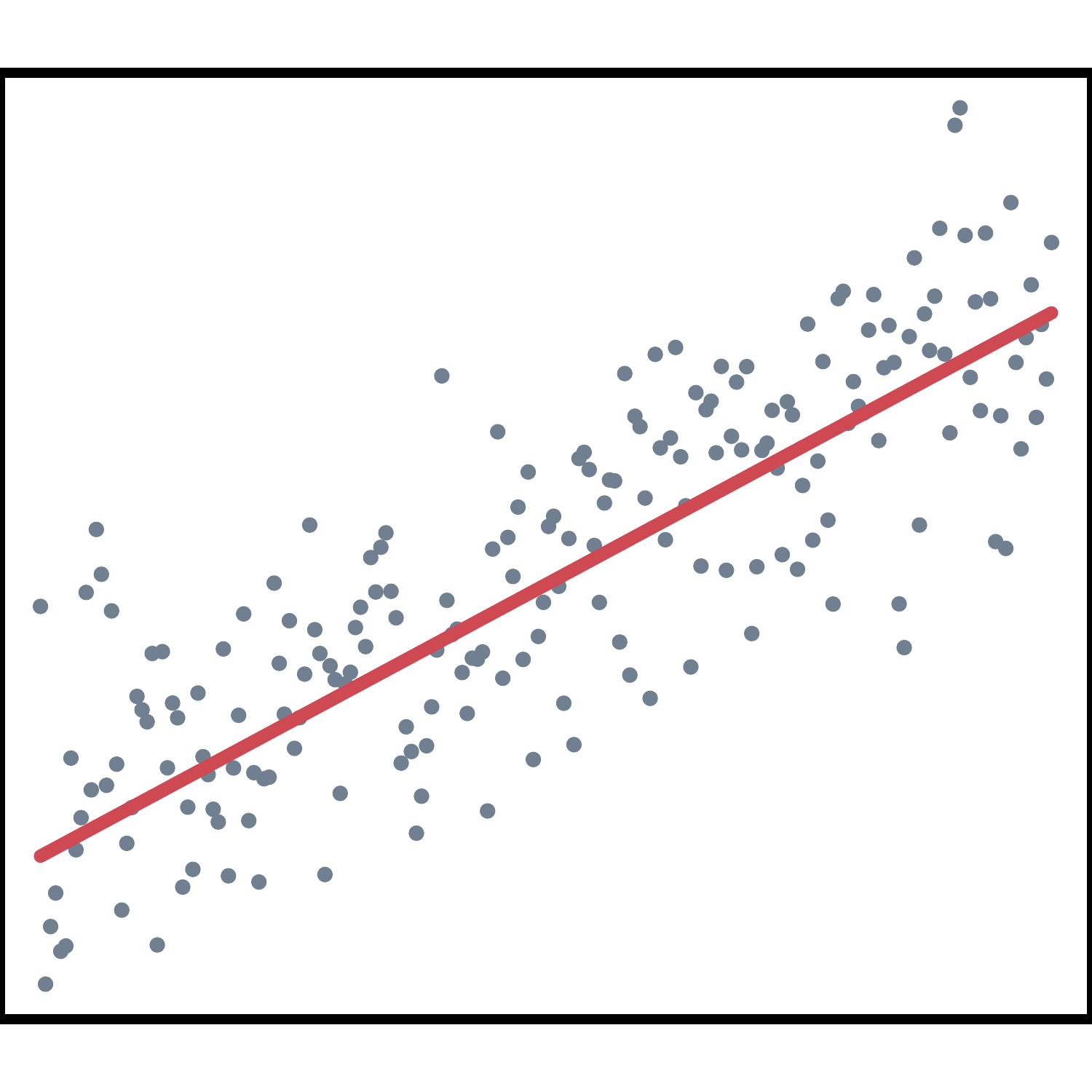}
\endminipage};
\node (b) at (5,0) 
{
\minipage{0.29\textwidth}
\centering

\includegraphics{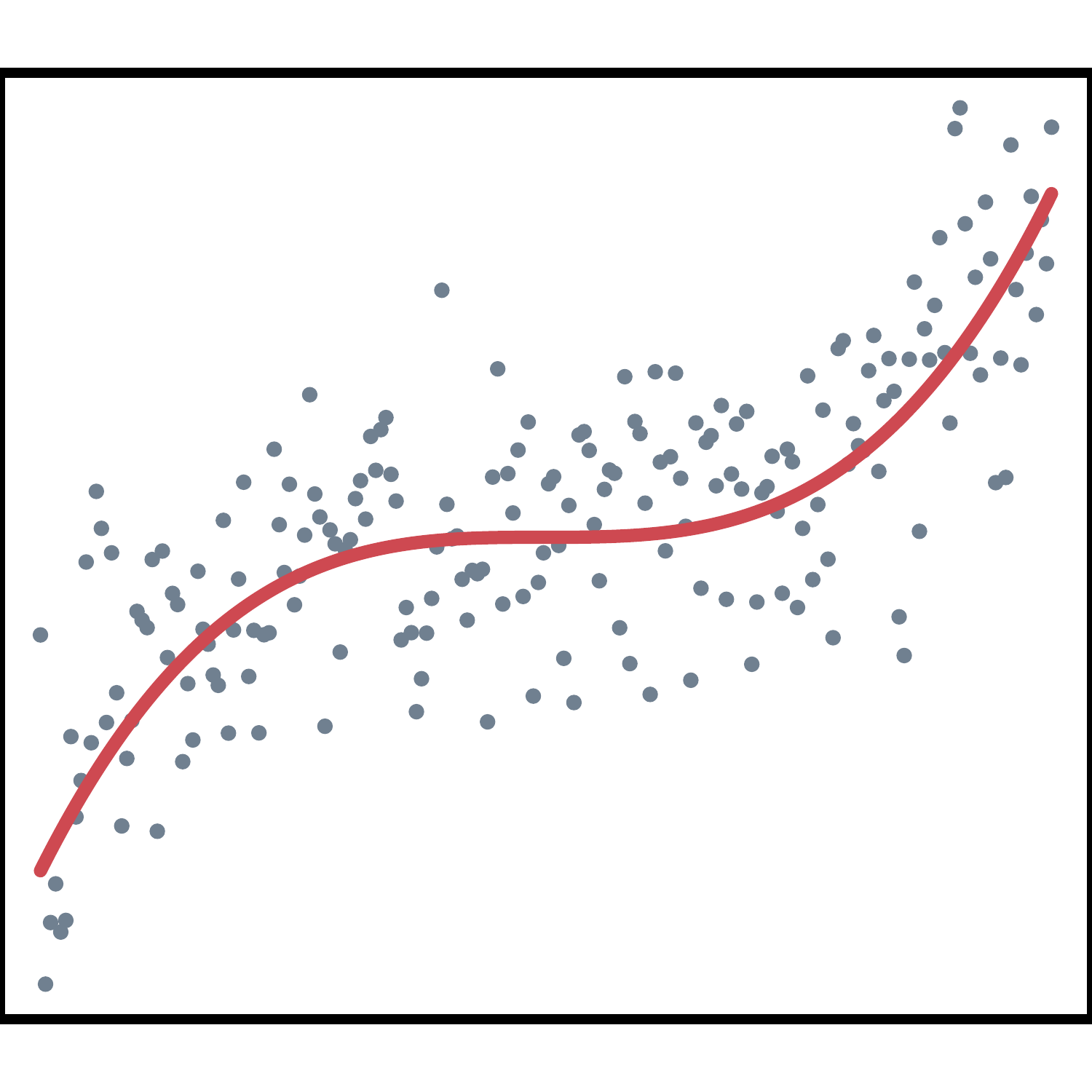}
\endminipage};
\node (c) at (10,0) 
{
\minipage{0.29\textwidth}
\centering

\includegraphics{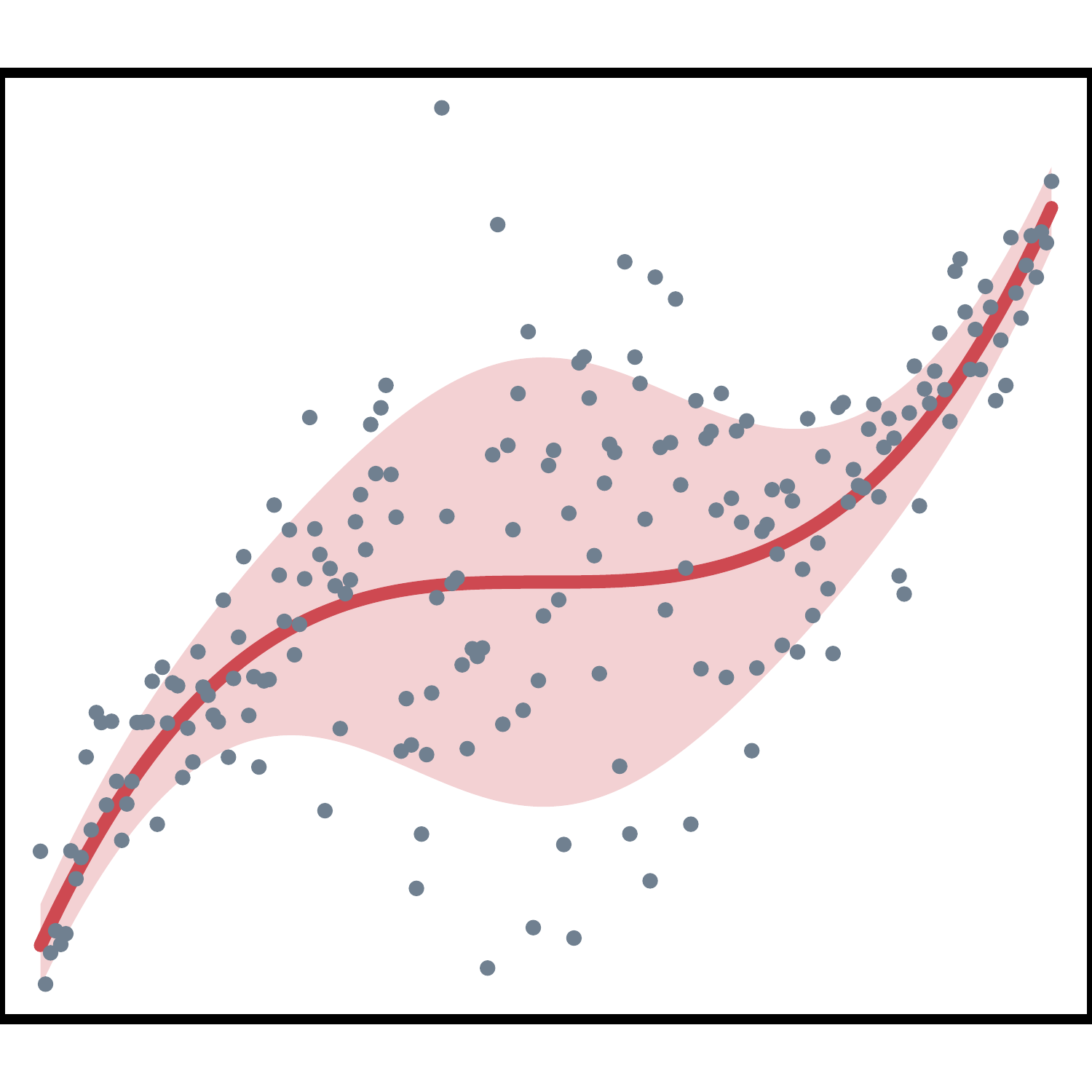}
\endminipage};
\node (d) at (0,-2) 
{
\minipage{0.29\textwidth}
\centering
\vspace{0.0cm}
LM, GLM  
\endminipage};
\node (e) at (5,-2) 
{
\minipage{0.29\textwidth}
\centering
\vspace{0.0cm}
GAM 
\endminipage};
\node at (10,-2)
{
\minipage{0.28\textwidth}
\centering
\vspace{0.0cm}
GAMLSS 
\endminipage};
\draw[-{>[scale=4, length=2, width=3]},line width=0.4pt](2.1,0)--(2.9,0);
\draw[-{>[scale=4, length=2, width=3]},line width=0.4pt](7.1,0)--(7.9,0);
\end{tikzpicture}
\caption{\label{fig:gam}Parametric modeling developments. (Generalized) linear models (left),
generalized additive models (middle), generalized additive models for location, scale, and shape (right).}
\end{figure}

In regression analysis a wide range of models has been developed to describe the
relationship between a response variable and a set of covariates. The classical
model is the linear model (LM) where the conditional mean of the response
is modeled through a linear function of the covariates (see the left panel of
Figure~\ref{fig:gam} for a schematic illustration). Over the last decades
this has been extended in various directions including:

\pagebreak

\begin{itemize}
  \item \emph{Generalized linear models} (GLMs, \citealp{Nelder+Wedderburn:1972})
    encompassing an additional nonlinear link function for the conditional mean.
  \item \emph{Generalized additive models} (GAMs, \citealp{Hastie+Tibshirani:1986})
    allowing for smooth nonlinear effects in the covariates
    (Figure~\ref{fig:gam}, middle).
  \item \emph{Generalized additive models for location, scale, and shape}
    (GAMLSS, \citealp{Rigby+Stasinopoulos:2005}) adopting a probabilistic modeling
    approach. In GAMLSS, each parameter of a statistical distribution can depend
    on an additive predictor of the covariates comprising linear and/or
    smooth nonlinear terms (Figure~\ref{fig:gam}, right).
\end{itemize}
Thus, the above-mentioned models provide a broad toolbox for capturing different
aspects of the response (mean only vs.\ full distribution) and different types
of dependencies on the covariates (linear vs.\ nonlinear additive terms).

While in many applications conditional mean regression models have been receiving
the most attention, there has been a paradigm shift over the last decade towards
distributional regression models. An important reason for this is that in many
fields forecasts of the mean are not the only (or not even the main) concern but
instead there is an increasing interest in probabilistic forecasts. Quantities of
interest typically include exceedance probabilities for certain thresholds of the
response or quantiles of the response distribution. Specifically, consider
weather forecasting where there is less interest in the mean amount of
precipitation on the next day. Instead, the probability of rain vs.\ no rain
is typically more relevant or, in some situations, a prediction interval of
expected precipitation (say from the expected 10\% to 90\% quantiles). Similar
considerations apply for other meteorological quantities and hence attention
in the weather forecasting literature has been shifting from classical linear 
deterministic models \citep{Glahn+Lowry:1972} towards probabilistic models such as the
non-homogeneous Gaussian regression (NGR) of \cite{Gneiting+Raftery+Westveld:2005}.
The NGR typically describes the mean of some meteorological response variable through
the average of the corresponding quantity from an ensemble of physically-based
numerical weather predictions (NWPs). Similarly, the variance of the response
is captured through the variance of the ensemble of NWPs. Thus, the NGR
considers both the mean as well as the uncertainty of the ensemble predictions
to obtain probabilistic forecasts calibrated to a particular site.

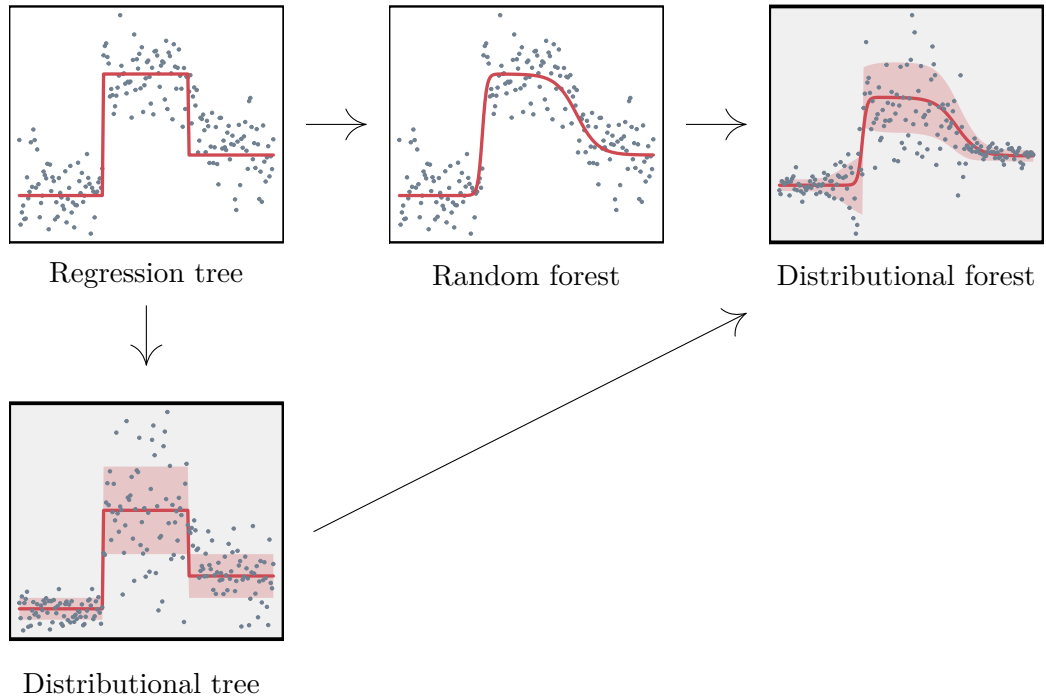
\begin{figure}[t!]
\begin{tikzpicture}
\node (a) at (0,0){
\minipage{0.29\textwidth}
\centering

\includegraphics{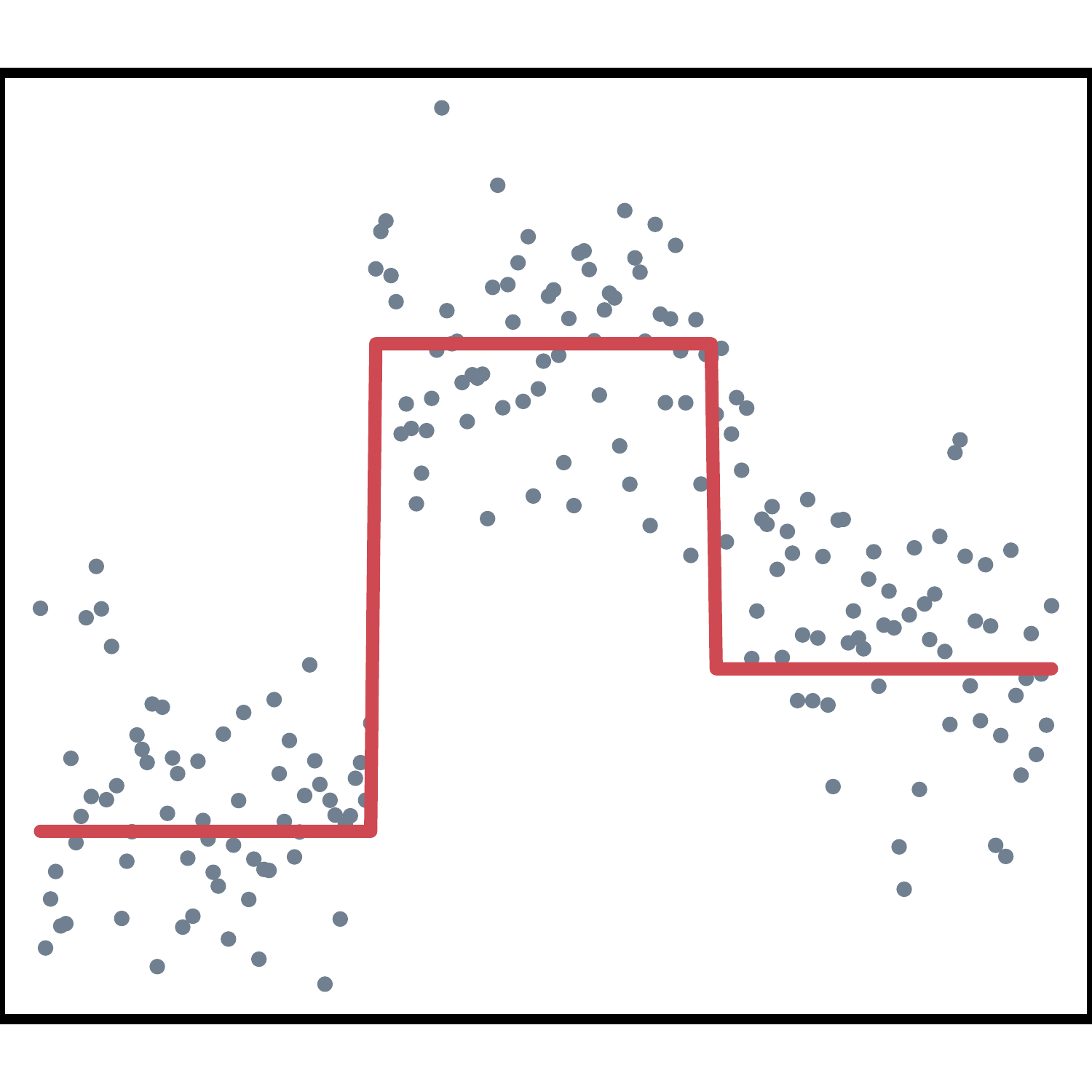}
\endminipage};
\node (b) at (5,0) 
{
\minipage{0.29\textwidth}
\centering

\includegraphics{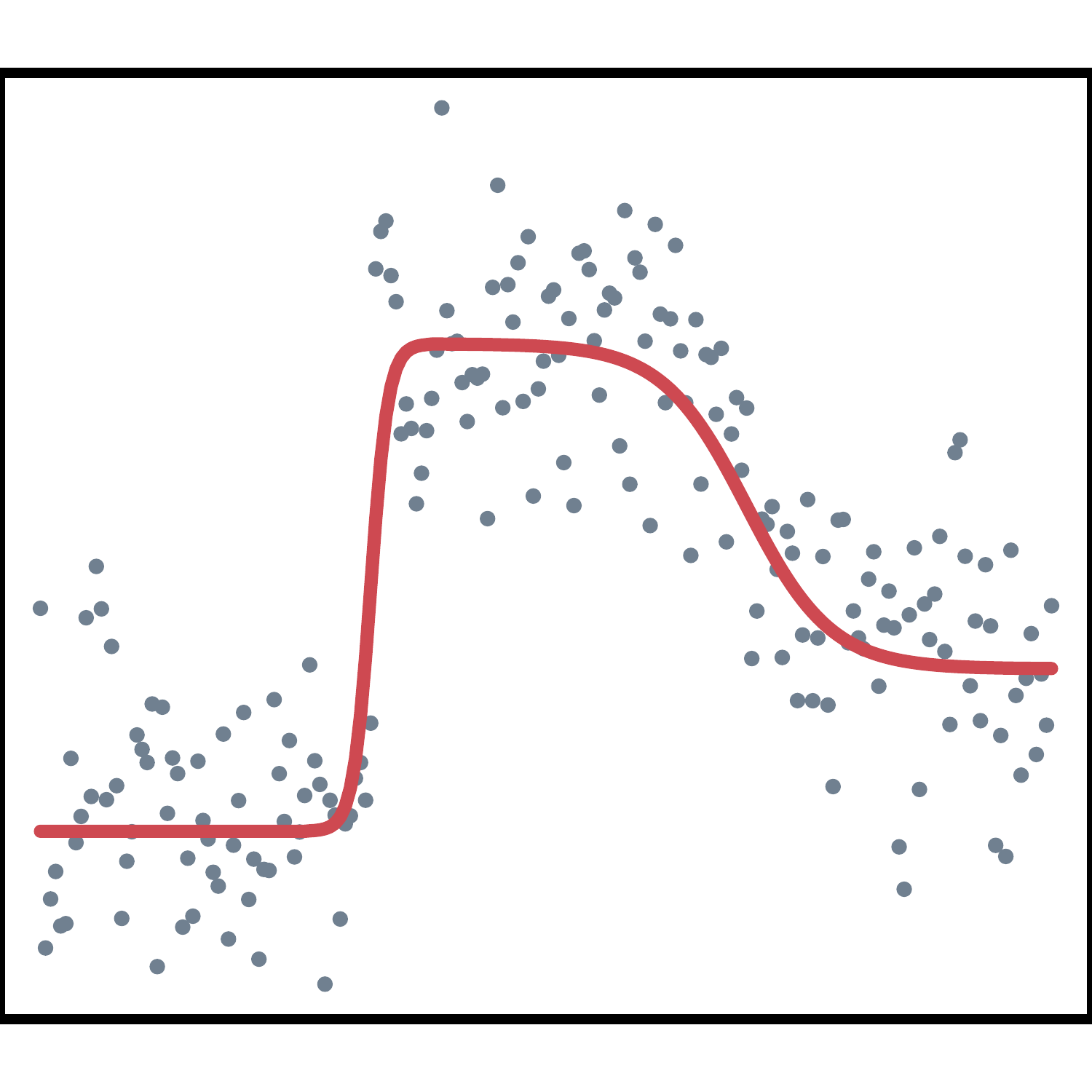}
\endminipage};
\node (c) at (10,0) 
{
\minipage{0.29\textwidth}
\centering

\includegraphics{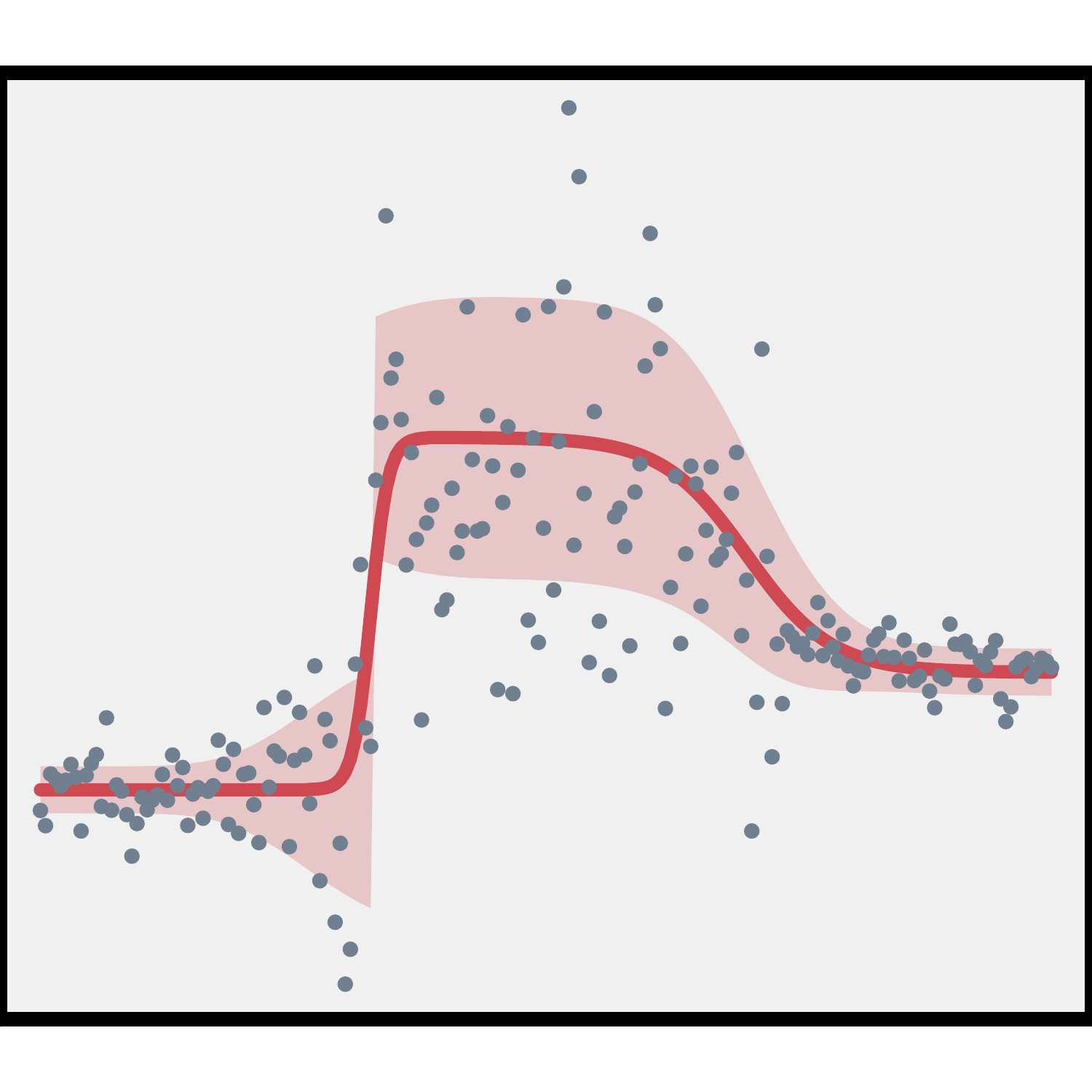}
\endminipage};
\node (d) at (0,-2) 
{
\minipage{0.29\textwidth}
\centering
\vspace{0.0cm}
Regression tree 
\endminipage};
\node (e) at (5,-2) 
{
\minipage{0.29\textwidth}
\centering
\vspace{0.0cm}
Random forest 
\endminipage};
\node at (10,-2)
{
\minipage{0.28\textwidth}
\centering
\vspace{0.0cm}
Distributional forest 
\endminipage};
\node (g) at (0,-5.4) 
{
\minipage{0.29\textwidth}
\vspace{0.2cm}
\centering

\includegraphics{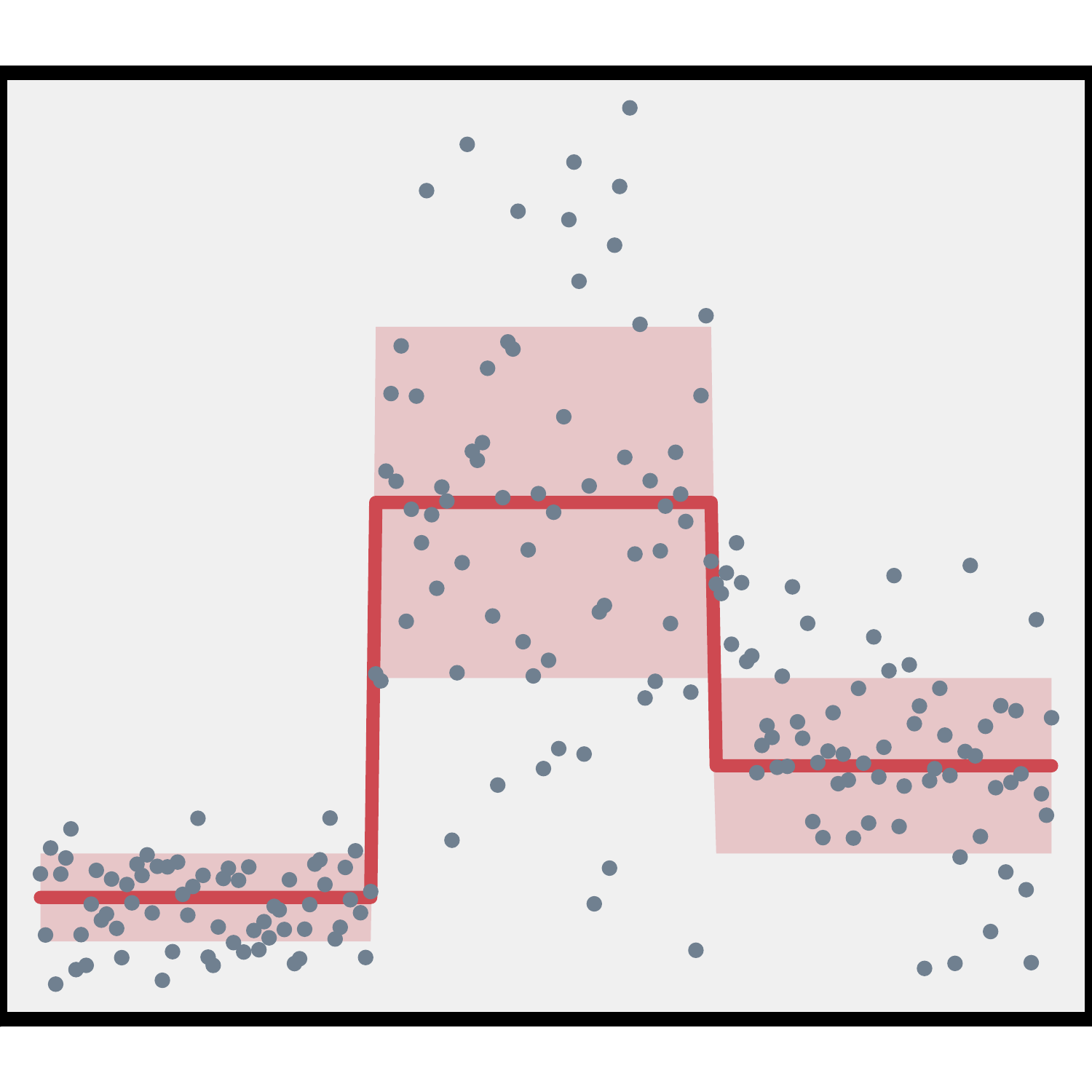}
\vspace{-0.2cm}
Distributional tree
\vspace{0.2cm}
\endminipage};
\draw[-{>[scale=4, length=2, width=3]},line width=0.4pt](2.1,0)--(2.9,0);
\draw[-{>[scale=4, length=2, width=3]},line width=0.4pt](7.1,0)--(7.9,0);
\draw[-{>[scale=4, length=2, width=3]},line width=0.4pt](0,-2.4)--(0,-3.2);
\draw[-{>[scale=4, length=2, width=3]},line width=0.4pt](2.2,-5.4)--(7.9,-2.5);
\end{tikzpicture}
\caption{\label{fig:tree}Tree and forest developments.
Regression tree (top left), distributional tree (bottom left), random forest
(top middle), and distributional forest (top right).}
\end{figure}

In summary, the models discussed so far provide a broad and powerful toolset for
parametric distributional fits depending on a specified set of additive linear or
smooth nonlinear terms. A rather different approach to capturing the dependence on
covariates are tree-based models.
\begin{itemize}
  \item \emph{Regression trees} (\citealp{Breiman+Friedman+Stone:1984}) recursively
    split the data into more homogeneous subgroups and can thus capture abrupt shifts
    (Figure~\ref{fig:tree}, top left) and approximate nonlinear functions.
    Furthermore, trees automatically carry out a forward selection of covariates and
    their interactions. 
  \item \emph{Random forests} (\citealp{Breiman:2001}) average the predictions of
    an ensemble of trees fitted to resampled versions of the learning data. This
    stabilizes the recursive partitions from individual trees and hence better
    approximates smooth functions (Figure~\ref{fig:tree}, top middle).
\end{itemize}
While classical regression trees and random forests only model the mean of the response
we propose to follow the ideas from GAMLSS modeling -- as outlined
in Figure~\ref{fig:gam} -- and combine tree-based methods with parametric distributional models,
yielding two novel techniques:
\begin{itemize}
  \item \emph{Distributional regression trees} (for short: \emph{distributional trees}) split the data into more homogeneous groups with
    respect to a parametric distribution, thus capturing changes in any distribution
    parameter like location, scale, or shape (Figure~\ref{fig:tree}, bottom left).
  \item \emph{Distributional regression forests} (for short: \emph{distributional forests}) utilize an ensemble of distributional trees
    for obtaining stabilized and smoothed parametric predictions (Figure~\ref{fig:tree},
    top right).
\end{itemize}

In the following, particular focus is given to distributional forests as a
method for obtaining probabilistic forecasts by leveraging the strengths of
random forests: the  ability to capture both smooth and abruptly changing
functions along with simultaneous selection of variables and possibly complex
interactions. Thus, these properties make the method particularly appealing in
case of many covariates with unknown effects and  interactions where it would be
challenging to specify a distributional regression model  like GAMLSS.
However, distributional forests should not be considered as a replacement
of GAMLSS but rather as a complementing technique for flexible distributional
regression -- much like GAMs and random forests are complements for conditional
mean regression.

\begin{figure}[t!]
\centering
\setkeys{Gin}{width=0.75\textwidth}
\includegraphics{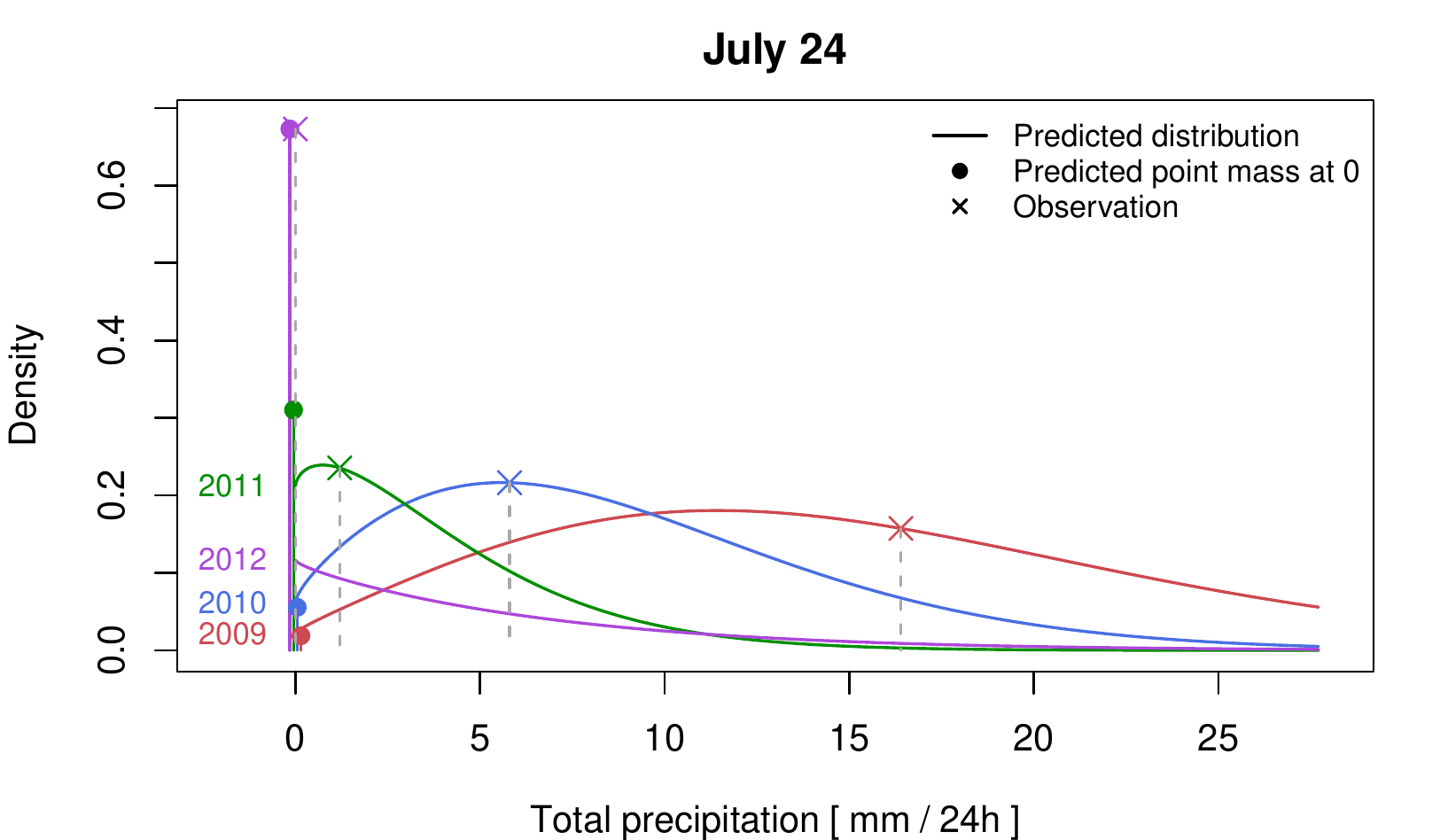}
\caption{\label{fig:axams-dist}Total precipitation predictions by a 
distributional forest at station Axams for July~24 in 2009, 2010, 2011 and 
2012 learned on data from 1985--2008. Observations are non-negative and 
modeled by a Gaussian distribution left-censored at zero.
The observations are depicted by crosses and the predicted point mass
from the model by filled circles.}
\end{figure}

In weather forecasting, the flexibility of distributional forests is especially 
appealing in mountainous regions and complex terrain where a wide range of local-scale
effects are not yet resolved by the NWP models. 
Thus, effects with abrupt changes and possibly nonlinear interactions 
might be required to account for site-specific unresolved features.
To illustrate this in practice,
precipitation forecasts are obtained with distributional forests at 95
meteorological stations in a mountainous region in the Alps, covering mainly
Tyrol, Austria, and adjacent areas (see the map in Figure~\ref{fig:all-map}).
More specifically, a Gaussian distribution left-censored at zero, is employed 
to model 24-hour total precipitation so that the
zero-censored point mass describes the probability of observing no precipitation 
on a given day (see Figure~\ref{fig:axams-dist}). Forecasts for July are
established based on data from the same month over the years 1985--2012 including
80~covariates derived from a wide range of different NWP quantities. As
Figure~\ref{fig:axams-dist} shows, the station-wise forests yield a full
distributional forecast for each day -- here for one specific day (July~24) at one
station (Axams) over four years (2009--2012) -- based on the previous 24~years
as learning data. The corresponding observations conform reasonably well with
the predictions. In Section~\ref{sec:precipitation} we investigate the
performance of distributional forests in this forecasting task in more detail.
Compared to three alternative zero-censored Gaussian models distributional forests
perform at least on par and sometimes clearly better while requiring no meteorological
knowledge about the atmospheric processes which drive formation of precipitation
for the model specification. The three alternatives are: a standard ensemble model
output statistics approach \citep[EMOS,][]{Gneiting+Raftery+Westveld:2005} based
on an NGR, a GAMLSS with regressors prespecified based on meteorological expertise
\citep[following][]{Stauffer+Umlauf+Messner:2017}, and a boosted GAMLSS
\citep{Hofner+Mayr+Schmid:2016} using non-homogeneous boosting
\citep{Messner+Mayr+Zeileis:2017} as an alternative technique for variable selection 
among all 80~available regressors.

\section{Methodology}
\label{sec:methods}

To embed the distributional approach from GAMLSS into regression trees and random
forests, we proceed in three steps. (1)~To fix notation, we briefly review fitting
distributions using standard maximum likelihood in Section~\ref{sec:distfit}.
(2)~A recursive partitioning strategy based on the corresponding scores (or
gradients) is introduced in Section~\ref{sec:disttree}, leading to distributional
trees. (3)~Ensembles of distributional trees fitted to randomized subsamples are
employed to establish distributional forests in Section~\ref{sec:distforest}.

The general distributional notation is exemplified in all three steps using the
Gaussian distribution left-censored at zero (for short: zero-censored Gaussian).
The latter is employed in the empirical case study in Section~\ref{sec:precipitation}
to model power-transformed daily precipitation amounts.

\subsection{Distributional fit}
\label{sec:distfit}

A distributional model $\mathcal{D}(Y, \bm{\theta})$ is considered for the response
variable $Y \in \mathcal{Y}$ using the distributional family $\mathcal{D}$ with
\textit{k}-dimensional parameter vector $\bm{\theta} \in \bm{\Theta}$ and
corresponding log-likelihood function $\ell(\bm{\theta}; Y)$. The GAMLSS
framework \citep{Rigby+Stasinopoulos:2005} provides a wide range of such
distributional families with parameterizations corresponding to location, scale,
and shape. Furthermore, censoring and/or truncation of these distributions
can be incorporated in the usual straightforward way
\citep[see e.g.,][Chapter~7.2]{Long:1997}.

To capture both location and scale of the probabilistic precipitation forecasts
while accounting for a point mass at zero (i.e., dry days without
rain), a zero-censored Gaussian distribution
with location parameter $\mu$ and scale parameter $\sigma$ is employed. Therefore,
the corresponding log-likelihood function with parameter vector 
$\bm{\theta} = (\mu, \sigma)$ is
\begin{equation}
\ell(\mu, \sigma; Y) = 
\begin{cases}
    \log\left\{\frac{1}{\sigma} \cdot \phi\left(\frac{Y - \mu}{\sigma}\right) \right\}, & \text{if } Y > 0\\[0.2cm]
    \log\left\{\Phi\left(\frac{-\mu}{\sigma}\right)\right\}, & \text{if } Y = 0
\end{cases}
\end{equation}

where $\phi$ and $\Phi$ are the probability density function and the cumulative
distribution function of the standard normal distribution $\mathcal{N}(0,1)$. 
Other distributions $\mathcal{D}$ and corresponding log-likelihoods
$\ell(\mu, \sigma; Y)$ could be set up in the same way, e.g., for
censored shifted gamma distributions \citep{Scheuerer+Hamill:2015} or
zero-censored logistic distributions \citep{Gebetsberger+Messner+Mayr:2017}.

With the specification of the distribution family and its log-likelihood 
function the task of fitting a distributional model turns into the task 
of estimating the distribution parameter~$\bm{\theta}$. This is commonly done by
maximum likelihood (ML) based on the learning sample with observations
$\{y_i\}_{i = 1, \dots, n}$ of the response variable $Y$. The maximum
likelihood estimator (MLE) $\bm{\hat \theta}$ is given by
\begin{equation}
\label{eq:mle}
\bm{\hat \theta} = \argmax_{\bm{\theta} \in \bm{\Theta}} \sum_{i=1}^n \ell(\bm{\theta}; y_i).
\end{equation}
Equivalently, this can be defined based on the corresponding first-order
conditions
\begin{equation}
\label{eq:foc}
\sum_{i = 1}^n s(\bm{\hat{\theta}}, y_i) = 0,
\end{equation}
where $s(\bm{\theta}; y_i)$ is the associated score function
\begin{equation}
\label{eq:score}
s(\bm{\theta}; y_i) = \frac{\partial \ell}{\partial \bm{\theta}}(\bm{\theta}; y_i).
\end{equation}
The latter is subsequently employed as a general goodness-of-fit measure
to assess how well the distribution with parameters $\bm{\theta}$ fits
one individual observation $y_i$.

\subsection{Distributional tree}
\label{sec:disttree}

Typically, a single global model $\mathcal{D}(Y, \bm{\theta})$ is not sufficient
for reasonably representing the response distribution. Therefore, covariates
$\bm{Z} = Z_1, \dots, Z_m \in \mathcal{Z}$ are
employed to capture differences in the distribution parameters $\bm{\theta}$.
In weather forecasting, these covariates typically include the output
from numerical weather prediction systems and/or lagged meteorological observations.

To incorporate the covariates into the distributional model, 
they are considered as regressors in additive predictors
$g_j(\theta_j) = f_{j,1}(\bm{Z}) + f_{j,2}(\bm{Z}) + \dots$
in GAMLSS.
Link functions $g_j(\cdot)$ are used for every parameter $\theta_j$
($j = 1, \dots, k$) based on smooth terms $f_{j,k}$ such as
nonlinear effects, spatial effects, random coefficients, or interaction surfaces
\citep{Klein+Kneib+Lang:2015}. However, this requires specifying the
additive terms and their functional forms in advance which can be challenging in
practice and potentially require expert knowledge in the application domain, 
especially if the number of covariates $m$ is large.

Regression trees generally take a different approach for automatically
including covariates in a data-driven way and allowing for abrupt
changes, nonlinear and non-additive effects, and interactions. In the context of
distributional models the goal is to partition the covariate space
$\mathcal{Z}$ recursively into disjoint segments so that a homogeneous
distributional model for the response $Y$ can be found for each segment with segment-specific
parameters. More specifically, the $B$ disjoint segments $\mathcal{B}_b$
($b = 1, \dots, B$) partition the covariate space
\begin{equation}
\label{eq:partition}
\mathcal{Z} = \dot{\bigcup\limits_{b = 1, \ldots, \textit{B}}} \mathcal{B}_b,
\end{equation}
and a local distributional model $\mathcal{D}(Y, \bm{\theta}^{(b)})$
(i.e., with segment-specific parameters $\bm{\theta}^{(b)}$) is fitted to the
response $Y$ in each segment.

To find the segments $\mathcal{B}_b$ that are (approximately) homogeneous
with respect to the distributional model with given parameters, the idea
is to use a gradient-based recursive-partitioning approach. In a given
subsample of the learning data this fits the model by ML (see
Equation~\ref{eq:mle}) and then assesses the goodness of fit by assessing the
corresponding scores $s(\bm{\hat \theta}; y_i)$ (see Equation~\ref{eq:score}).

To sum up, distributional trees are fitted recursively via:
\begin{enumerate}
\item Estimate $\bm{\hat \theta}$ via maximum likelihood for the observations
  in the current subsample.
\item Test for associations (or instabilities) of the scores $s(\bm{\hat \theta}, y_i)$ 
  and $Z_{l,i}$ for each partitioning variable~$Z_l$ ($l = 1, \dots, m$).
\item Split the sample along the partitioning variable $Z_l^*$ with the
  strongest association or instability. Choose the breakpoint with the highest
  improvement in the log-likelihood or the highest discrepancy.
\item Repeat steps 1--3 recursively in the subsamples until these become too
  small or there is no significant association/instability (or some other
  stopping criterion is reached).
\end{enumerate}
Different inference techniques can be used for assessing the association between
scores and covariates in step~3. In the following we use the general class of
permutation tests introduced by \cite{Hothorn+Hornik+VanDeWiel:2006} which is
also the basis of conditional inference trees \citep[CTree,][]{Hothorn+Hornik+Zeileis:2006}.
Alternatively, one could use asymptotic M-fluctuation tests for parameter
instability \citep{Zeileis+Hornik:2007} as in model-based recursive partitioning
\citep[MOB,][]{Zeileis+Hothorn+Hornik:2008}. More details are provided in
Appendix~\ref{app:tree}.

For obtaining probabilistic predictions from the tree for a (possibly new) set
of covariates $\bm{z} = (z_1, \ldots, z_m)$, the observation simply has to
be ``sent down'' the tree and the corresponding segment-specific MLE has to be
obtained. Thus, in practice $\bm{\hat \theta}(\bm{z})$ does not have to be 
recalculated for each new $\bm{z}$ but one can simply extract the parameter estimates 
for the corresponding segment which have been computed already while
learning the tree. However, to understand this estimator conceptually it is useful
to denote it as a weighted MLE where the weights select those observations from 
the learning sample that fall into the same segment:
\begin{equation}
w^{\text{tree}}_i(\bm{z}) = \sum_{b=1}^B \mathbf{1}((\bm{z}_i \in \mathcal{B}_b) \land (\bm{z} \in \mathcal{B}_b)),
\end{equation}
where $\mathbf{1}(\cdot)$ is the indicator function. The predicted distribution
for a given $\bm{z}$ is then fully specified by the estimated parameter
$\bm{\hat \theta}(\bm{z})$ where
\begin{equation}
\bm{\hat \theta}(\bm{z}) = \argmax_{\bm{\theta} \in \bm{\Theta}} \sum_{i=1}^n w^{\text{tree}}_i(\bm{z}) \cdot \ell(\bm{\theta}; y_i).
\end{equation}

\subsection{Distributional forest}
\label{sec:distforest}

While the simple recursive structure of a tree model is easy to visualize and
interpret, the abrupt changes are often too rough, instable, and impose steps
on the model even if the true underlying effect is smooth. Hence, ensemble
methods such as bagging or random forests \citep{Breiman:2001} are typically
applied to smooth the effects, stabilize the model, and improve predictive performance.

The idea of random forests is to learn an ensemble of trees, each on a different
learning data obtained through resampling (bootstrap or subsampling). In each node 
only a random subset of the covariates $\bm{Z}$ is considered for splitting to reduce the 
correlation among the trees and to stabilize the variance of the model.
For a simple regression random forest the mean of predictions over
all trees is considered. In that way changes in the location  of the response 
across the covariates are detected (e.g., in Breiman and Cutler's random forests, 
\citealp{Breiman:2001}).
This idea is now taken one step further by embedding it in a distributional 
framework based on maximum-likelihood estimation. \emph{Distributional forests} 
employ an ensemble of $T$~\emph{distributional trees} which pick up changes in 
the ``direction'' of any distribution parameter by considering the full likelihood
and corresponding score function for choosing splitting variables and split points.

To obtain probabilistic predictions from a distributional forest, it still
needs to be specified how to compute the parameter estimates
$\bm{\hat \theta}(\bm{z})$ for a (potentially new) set of covariates $\bm{z}$.
Following \cite{Hothorn+Zeileis:2017} we interpret random forests as adaptive
local likelihood estimators using the averaged ``nearest neighbor weights''
\citep{Lin+Jeon:2006} from the $T$~trees in the forest
\begin{equation}
w^{\text{forest}}_i(\bm{z}) = \frac{1}{T} \sum_{t=1}^T \sum_{b=1}^{B^t}
\frac{\mathbf{1}((\bm{z}_i \in \mathcal{B}^t_b) \land (\bm{z} \in \mathcal{B}^t_b))}{|\mathcal{B}^t_b|},
\end{equation}
where $|\mathcal{B}^t_b|$ denotes the number of observations in the $b$-th
segment of the $t$-th tree.
Thus, these $w^{\text{forest}}_i(\bm{z}) \in [0, 1]$ whereas
$w^{\text{tree}}_i(\bm{z}) \in \{0, 1\}$. Hence, weights cannot only be $0$
or $1$ but change more smoothly, giving high weight to those observations $i$
from the learning sample that co-occur in the same segment $\mathcal{B}_b^t$
as the new observation $\bm{z}$ for many of the trees $t = 1, \dots, T$.
Consequently, the parameter estimates may, in principle, change for every
observation and can be obtained by
\begin{equation}
\bm{\hat \theta}(\bm{z}) =  \argmax_{\bm{\theta} \in \bm{\Theta}} \sum_{i=1}^n w^{\text{forest}}_i(\bm{z}) \cdot \ell(\bm{\theta}; y_i).
\end{equation}
In summary, this yields a parametric distributional regression model 
(through the score-based approach) that can capture both abrupt effects and
high-order interactions (through the trees) and smooth effects (through
the forest).

Distributional forests share some concepts and algorithmic aspects 
with other generalizations of Breiman and Cutler's random forests. Nearest 
neighbor weights are employed for aggregation in survival forests 
\citep{Hothorn+Lausen+Benner:2004},
quantile regression forests \citep{Meinshausen:2006},
transformation forests \citep{Hothorn+Zeileis:2017},
and generalized random forests for causal inferences 
\citep{Athey+Tibshirani+Wager:2019}.
These procedures aggregate over trees fitted to specific score functions 
(e.g., log rank scores in survival trees, model residuals in transformation 
or generalized forests). Distributional forests, in contrast to 
nonparametric approaches, provide a compromise between model flexibility 
and interpretability: The parameters of a problem-specific distribution 
(zero-censored Gaussian for precipitation) have a clear meaning but 
may depend on external variables in a quite general way.

\section{Probabilistic precipitation forecasting in complex terrain}
\label{sec:precipitation}

Many statistical weather forecasting models leverage the strengths of modern
numerical ensemble prediction systems (EPSs; see \citealp{Bauer+Thorpe+Brunet:2015}).
EPSs not only predict the most likely future state of the atmosphere but provide
information about the uncertainty for a specific quantity and weather situation.
This is done by running the NWP model several times using slightly perturbed 
initial conditions and model specifications to account for uncertainties in both,
the initial atmospheric state and the NWP model (and its parametrizations).
One frequently-used method based on distributional regression models is the 
ensemble model output statistics (EMOS) approach first proposed by
\cite{Gneiting+Raftery+Westveld:2005} to produce high-quality forecasts for
specific quantities and sites.  
In case of precipitation forecasting, EMOS typically uses the ensemble mean of 
``total precipitation'' (\emph{tp}) forecasts as the predictor for the location 
parameter $\mu$ and the corresponding ensemble standard deviation for the 
scale parameter $\sigma$, e.g., assuming the observations to follow a zero-censored
Gaussian distribution.
This distributional approach of modeling both parameters allows to correct 
for possible errors of the NWP ensemble in both, the expectation but also the 
uncertainty of a specific forecast. Thus, a basic EMOS specification typically
models the two distribution parameters by two linear predictors, e.g., 
$\mu = \beta_0 + \beta_1 \cdot \mathrm{mean}(\emph{tp})$ and  
$\log(\sigma) = \gamma_0 + \gamma_1 \cdot \log(\mathrm{sd}(\emph{tp}))$
with regression coefficients $\beta_0$, $\beta_1$, $\gamma_0$, and $\gamma_1$
\citep[where the log link assures positivity of the scale parameter, following][]{Gebetsberger+Messner+Mayr:2017}.

%

While this approach alone is already highly effective in the plains,
it typically does not perform as well in complex terrain due to unresolved
effects in the NWP system \citep{Bauer+Thorpe+Brunet:2015}. 
For example, in the Tyrolean Alps -- considered
in the following case study -- the NWP grid cells of $50 \times 50$
km$^2$ are too coarse to capture single mountains, narrow valleys, etc. 
Therefore, it is often possible to substantially improve the predictive 
performance of a basic EMOS by including additional predictor variables, either 
from local meteorological observations or an NWP model. Unfortunately, it is 
typically unknown which variables are relevant for improving the predictions. 
Simply including all available variables may be computationally burdensome and
can lead to overfitting but, on the other hand, excluding too many variables
may result in a loss of valuable information. Therefore, selecting
the relevant variables and interactions among all possible covariates is 
crucial for improving the statistical forecasting model.

In the following, it is illustrated how distributional forests can
solve this problem without requiring prior expert knowledge about the 
meteorological covariates.
For fitting the forest only the response distribution and the list of potential predictor
variables need to be specified (along with a few algorithmic details) and then
the relevant variables, interactions, and potentially nonlinear effects are
determined automatically in a data-driven way. Here, we employ a zero-censored
Gaussian distribution and 80~predictor variables computed from ensemble means
and spreads of various NWP outputs. The predictive performance of the forest is
compared to three other zero-censored Gaussian models: (a)~a basic
EMOS, (b)~a GAMLSS with prespecified effects and interactions based on
meteorological knowledge/experience, and (c)~a boosted GAMLSS with automatic
selection of smooth additive terms based on all 80~predictor variables.

\subsection{Data}
\label{sec:data}

\begin{table}[t!]
\begin{minipage}{\textwidth}
\caption[Table caption text]{Basic covariates together with the number ({\#}) 
and the type of variations.
Time periods indicate aggregation time periods in hours after NWP model
initialization (e.g., 6--30 corresponds to +6\,h to +30\,h ahead forecasts,
0600\,UTC to 0600\,UTC of the next day). Note: $^*$Minimum values of \emph{dswrf}
over 24\,h are always zero and thus neglected.}
\label{tab:covariates}
\begin{tabular}{ l  c  l }
\hline
Basic covariates & {\#} & Variations\\
\hline
\emph{tp}: total precipitation,           & 12  & ensemble mean of sums over 24h, \\
\hspace*{0.5cm} power transformed (by $1.6^{-1}$) &    & ensemble std.\ deviation of sums over 24h, \\ 
\emph{cape}: convective available     &    & ensemble minimum of sums over 24h, \\
\hspace*{0.9cm} potential energy, &    & ensemble maximum of sums over 24h\\
\hspace*{0.9cm} power transformed (by $1.6^{-1}$)&    & \qquad all for 6--30 \\
\hspace*{\fill}                            &    & ensemble mean of sums over 6h\\
\hspace*{\fill}                            &    & \qquad for 6--12, 12--18, 18--24, 24--30 \\
\hspace*{\fill}                            &    & ensemble std.\ deviation of sums over 6h\\
\hspace*{\fill}                            &    & \qquad for 6--12, 12--18, 18--24, 24--30 \\
\hline
\emph{dswrf}: downwards short wave      & 6 & ensemble mean of mean values, \\
\hspace*{1.13cm} radiation flux (``sunshine'') &   & ensemble mean of minimum values$^*$,\\
\emph{msl}: mean sea level pressure     &   &  ensemble mean of maximal values,\\
\emph{pwat}: precipitable water         &   & ensemble std.\ deviation of mean values,\\
\emph{tmax}: 2m maximum temperature     &   & ensemble std.\ deviation of minimum values$^*$,\\
\hspace*{1.15cm}                          &   & ensemble std.\ deviation of maximal values,\\
\emph{tcolc}: total column-integrated   &   & \qquad all over 6--30 \\
\hspace*{0.95cm} condensate               &   & \\
\emph{t500}: temperature on 500 hPa     &   & \\
\hspace*{1cm}                             &   & \\
\emph{t700}: temperature on 700 hPa     &   & \\
\hspace*{1cm}                             &   & \\
\emph{t850}: temperature on 850 hPa     &   & \\
\hspace{1cm}                              &   & \\
\hline
\emph{tdiff500850}: temperature         & 3 & ensemble mean of difference in mean,\\
\hspace*{\fill} difference 500 to 850 hPa &   & ensemble minimum of difference in mean,\\
\emph{tdiff500700}: temperature         &   & ensemble maximum of difference in mean\\
\hspace*{\fill} difference 500 to 700 hPa &   & \qquad all over 6--30 \\
\emph{tdiff700850}: temperature         &   & \\
\hspace*{\fill} difference 700 to 850 hPa &   & \\
\hline
\emph{msl{\_}diff}: mean sea level pressure & 1 & \emph{msl{\_}mean{\_}max} $-$ \emph{msl{\_}mean{\_}min}\\
\hspace{1.6cm} difference            &   & \qquad over 6--30 \\
\hline
\end{tabular}
\end{minipage} 
\end{table}

Learning and validation data consist of observed 
daily precipitation sums provided by the National Hydrographical Service 
(\citealp{ehyd}) and numerical weather forecasts from the U.S.~National
Oceanic and Atmospheric Administration (NOAA).
Both, observations and forecasts are available for 1985--2012 and
the analysis is exemplified using July, the month with the most precipitation
in Tyrol.

Observations are obtained for 95~stations all over Tyrol and surroundings,
providing 24-hour precipitation sums measured at 0600\,UTC and rigorously
quality-checked by the National Hydrographical Service. NWP outputs
are obtained from the second-generation reforecast data set of the
global ensemble forecast system \citep[GEFS,][]{Hamill+Bates+Whitaker:2013}. 
This data set consists of an 11-member ensemble based on a fixed version of the 
numerical model and a horizontal grid spacing of about {$50 \times 50$ km$^2$}
initialized daily at 0000~UTC from December 1984 to present 
providing forecasts on a 6-hourly temporal resolution. Each of the 
11 ensemble members uses slightly different perturbed initial 
conditions to predict the situation-specific uncertainty 
of the atmospheric state. 

From the GEFS, 14 basic forecast variables are considered with up 
to 12 variations each such as mean/maximum/minimum
over different aggregation time periods. A detailed overview is
provided in Table~\ref{tab:covariates}, yielding 80~predictor variables in total.

To remove large parts of the skewness of precipitation data, 
a power transformation \citep{Box+Cox:1964} is often applied, e.g., using
cubic \citep{Stidd:1973} or square root \citep{Hutchinson:1998} transformations.
However, the power parameter may vary for different climatic zones or temporal 
aggregation periods and hence we follow \cite{Stauffer+Mayr+Messner:2017} 
in their choice of $1.6^{-1}$ as a suitable power parameter for precipitation
in the region of Tyrol. The same power transformation is applied to both the
observed precipitation sums and the NWP outputs ``total precipitation'' (\emph{tp}) and
``convective available potential energy'' (\emph{cape}).

\subsection{Models and evaluation}
\label{sec:evaluation}

\begin{table}[t!]
\centering
\caption[Table caption text]{Overview of models with type of 
covariate dependency and included covariates for each distribution 
parameter. \emph{A} $\ast$ \emph{B} indicates an interaction between
covariate \emph{A} and \emph{B}.}
\label{tab:models}
\begin{tabular}{ l l l l }
\hline
Model & Type & Location ($\mu$) & Scale ($\log(\sigma)$)                                  \\ \hline
Distributional forest & recursive    & all                & all                           \\ 
                      & partitioning &                    &                               \\ \hline
EMOS                  & linear       & \emph{tp{\_}mean}  & \emph{tp{\_}sprd}             \\ \hline
Prespecified GAMLSS   & spline       & \emph{tp{\_}mean}, & \emph{tp{\_}sprd},            \\
                      & in each      & \emph{tp{\_}max},  & \emph{dswrf{\_}sprd{\_}mean}, \\
 & & \emph{tp{\_}mean1218} $\ast$ & \emph{tp{\_}sprd1218} $\ast$\\ 
 & & \quad \emph{cape{\_}mean1218}, & \quad \emph{cape{\_}mean1218},\\
 & & \emph{dswrf{\_}mean{\_}mean}, & \emph{tcolc{\_}sprd{\_}mean},\\
 & & \emph{tcolc{\_}mean{\_}mean}, & \emph{tdiff500850{\_}mean}\\
 & & \emph{pwat{\_}mean{\_}mean}, & \\
 & & \emph{tdiff500850{\_}mean}, & \\
 & & \emph{msl{\_}diff} & \\ \hline
Boosted GAMLSS        & spline  & all & all \\
                      & in each &     &     \\ \hline
\end{tabular}
\end{table}

The following zero-censored Gaussian regression models are employed
in the empirical case study, see Table~\ref{tab:models} for
further details:
\begin{itemize}

\item \emph{Distributional forest:} All 80~predictor variables are
  considered for learning a forest of 100 trees. Subsampling is employed 
  for each tree using a third of the predictors in each split of the tree
  (argument \code{mtry} in our implementation \code{distforest},
  with more ``computational details'' provided at the end of the manuscript).
  Parameters are estimated by
  adaptive local likelihood based on the forest weights as described
  in Section~\ref{sec:methods}. 
  The stopping criteria are the minimal number of observations to perform a split 
  (\texttt{minsplit}~$= 50$), the minimal number of observations in a 
  segment (\texttt{minbucket}~$= 20$), and the significance level 
  for variable selection (\texttt{alpha}~$= 1$). The latter means
  that no early stopping (or ``pre-pruning'') is applied
  based on results of the statistical tests.
  
  

\item \emph{EMOS:} The basic ensemble model output statistics
  models use the ensemble mean of total precipitation as regressor in the location
  submodel and the corresponding ensemble standard deviation in the scale submodel.
  The parameters are estimated by maximum likelihood, using an identity
  link for the location part and a log link for the scale part
  \citep[following the advice of][]{Gebetsberger+Messner+Mayr:2017}.

\item \emph{Prespecified GAMLSS:} Smooth additive splines are selected for
  the most relevant predictors based on meteorological expert knowledge
  following \cite{Stauffer+Umlauf+Messner:2017}. More specifically, based
  on the 80~available variables, eight terms are included in the location
  submodel and five in the scale submodel. Both involve an interaction of
  \emph{tp} and \emph{cape} in the afternoon (between 1200\,UTC and 1800\,UTC)
  to capture the potential for thunderstorms that frequently occur in
  summer afternoons in the Alps. The model is estimated by maximum
  penalized likelihood using a backfitting algorithm \citep{Stasinopoulos+Rigby:2007}.

\item \emph{Boosted GAMLSS:} Smooth additive splines are selected
  automatically from all 80~available variables, using non-cyclic boosting
  for parameter estimation \citep{Hofner+Mayr+Schmid:2016,Messner+Mayr+Zeileis:2017}.
  This updates the predictor terms for the location or scale submodels iteratively
  by maximizing the log-likelihood only for the variable yielding the
  biggest improvement. The iteration stops early -- before fully maximizing
  the in-sample likelihood -- based on a (computationally intensive)
  out-of-bag bootstrap estimate of the log-likelihood. The grid considered for
  the number of boosting iterations (\code{mstop}) is: $50, 75, \dots, 975, 1000$.

\end{itemize}

The predictive performance in terms of full probabilistic forecasts is
assessed using the continuous ranked probability score (CRPS, \citealp{Hersbach:2000}).
For each of the models this assesses the discrepancy of the
predicted distribution function $F$ from the observation $y$ by
\begin{equation}
\text{CRPS}(y, F) = \int_{-\infty}^\infty (F(z) - \mathbf{1}(y \leq z))^2 dz
\end{equation}
where \(\mathbf{1}(\cdot)\) is the indicator function. In the subsequent
applications, the mean CRPS is always evaluated out of sample, 
either using cross-validation or a hold-out data set (2009--2012)
that was not used for learning (1985--2008). CRPS is a proper scoring rule
\citep{Gneiting+Raftery:2007} often used within the meteorological community.
Lower values indicate better performance.

To assess differences in the improvement of the forests and GAMLSS
models over the basic EMOS, a CRPS-based skill score with EMOS as
the reference method is computed:
\begin{equation}
\text{CRPSS}_{\text{method}} = 1 - \frac{\text{CRPS}_{\text{method}}}{\text{CRPS}_{\text{EMOS}}}.
\end{equation}

\subsection{Application for one station}
\label{sec:axams}

\begin{figure}[t!]
\centering
\setkeys{Gin}{width=0.99\textwidth}
\includegraphics{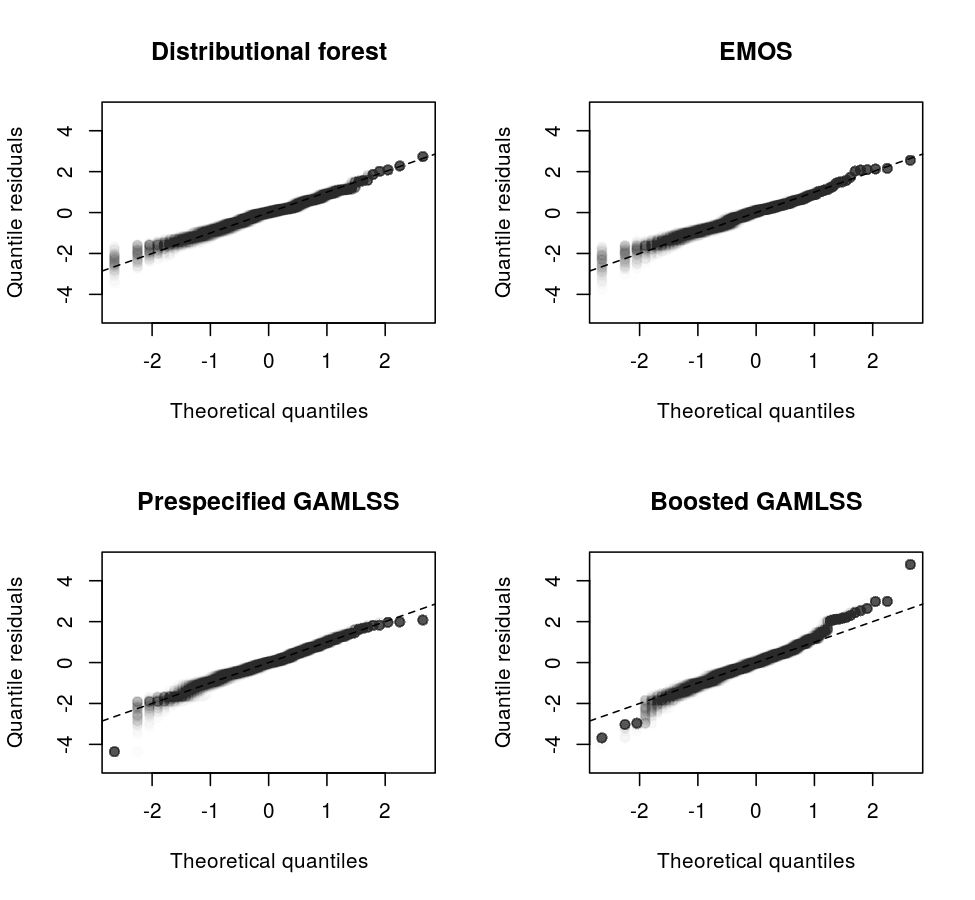}
\caption{\label{fig:axams-qq} Out-of-sample residual QQ~plots (2009--2012) 
for station Axams based on models learned on data from 1985--2008.}
\end{figure}

In a first step, we show a detailed comparison of the competing models
for one observation site, Axams in Tyrol (in the center of the study area,
see Figure~\ref{fig:all-map}). As for all other stations, daily precipitation 
observations and numerical weather predictions are available for the month of July 
from 1985 through 2012. In Figure~\ref{fig:axams-dist} in the introduction the 
probabilistic forecasts from the distributional forest, trained on 1985--2008, 
for July 24 in 2009--2012 have already been shown as a motivational example. 
In particular, the figure depicts the forecasted point mass at zero
(i.e., the probability of a dry day) along with the forecasted probability 
density function for the total amount of precipitation. Based on this illustration 
it can be observed that the four sample forecasts differ considerably in 
location $\mu$, scale $\sigma$, and the amount of censoring while conforming quite 
well with the actual observations from these days. 
While this is a nice illustrative example we are  
interested in the overall predictive performance and calibration of the
distributional fits. More details of this assessment as well as
an application to 14~further meteorological stations is provided in
Supplement~B \citep{Schlosser+Hothorn+Stauffer:2019b}.

To assess calibration, Figure~\ref{fig:axams-qq} shows residual QQ~plots for
out-of-sample predictions (2009--2012) from the different models trained on
1985--2008. Due to the point masses at zero 100 draws from the randomized
quantile residuals \citep{Dunn+Smyth:1996} are plotted in semi-transparent
gray. Overall, the randomized quantile residuals conform quite well with the
theoretical standard normal quantile (i.e., form a straight line close to the 
diagonal), indicating that all four models are sufficiently well calibrated. 
This is also supported by the corresponding probability integral transform
(PIT, \citealp{Gneiting+Balabdaoui+Raftery:2007}) histograms for station
Axams in Supplement~B \citep{Schlosser+Hothorn+Stauffer:2019b} which 
contains a more detailed explanation of residual QQ plots and PIT histograms.

\begin{figure}[t!]
\centering
\includegraphics{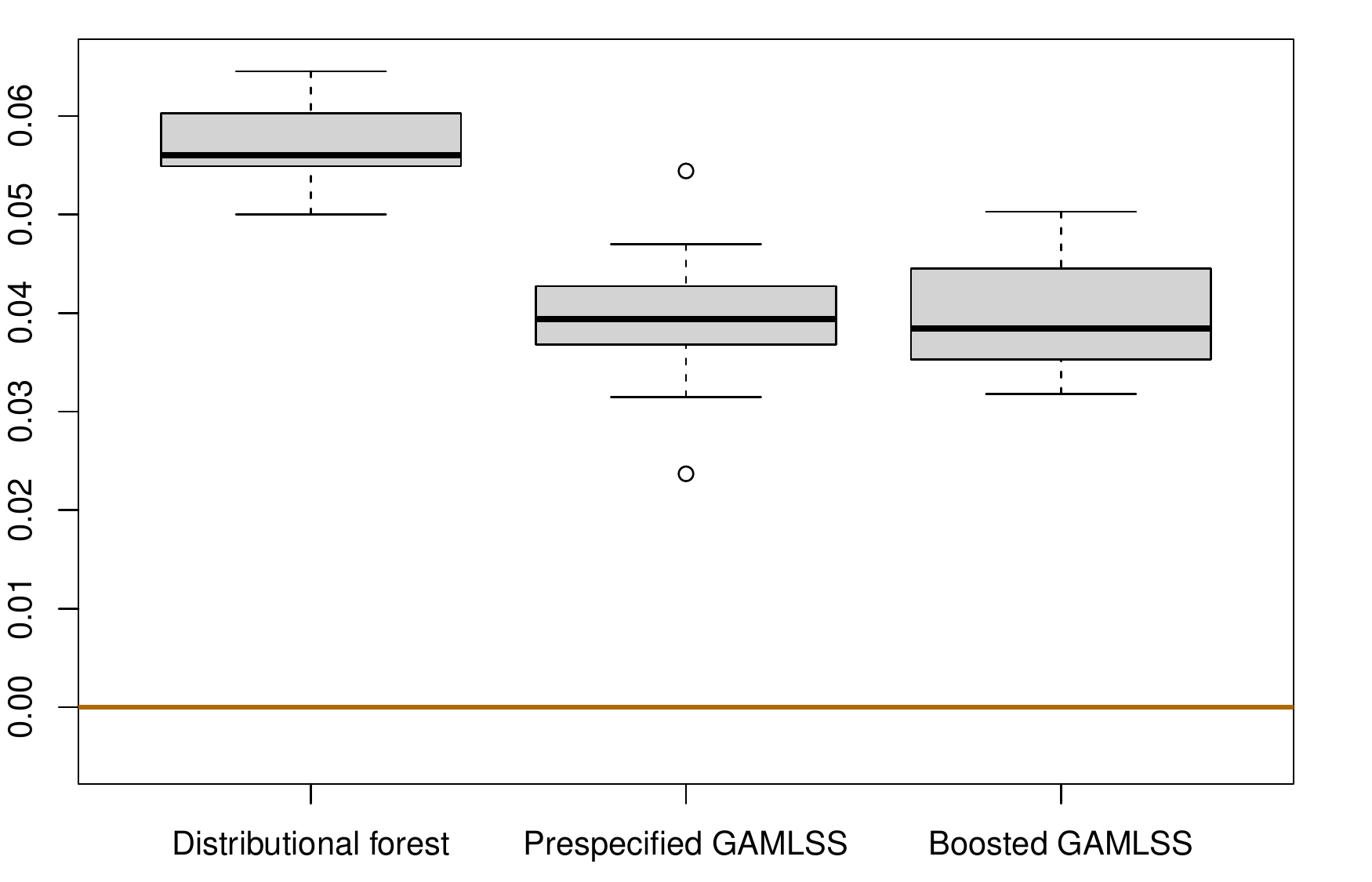}
\caption{\label{fig:axams-crps}CRPS skill score from the 10 times 
7-fold cross-validation at station Axams (1985--2012). The horizontal orange 
line pertains to the reference model EMOS.}
\end{figure}

\begin{figure}[t!]
\centering
\setkeys{Gin}{width=0.7\textwidth}
\includegraphics{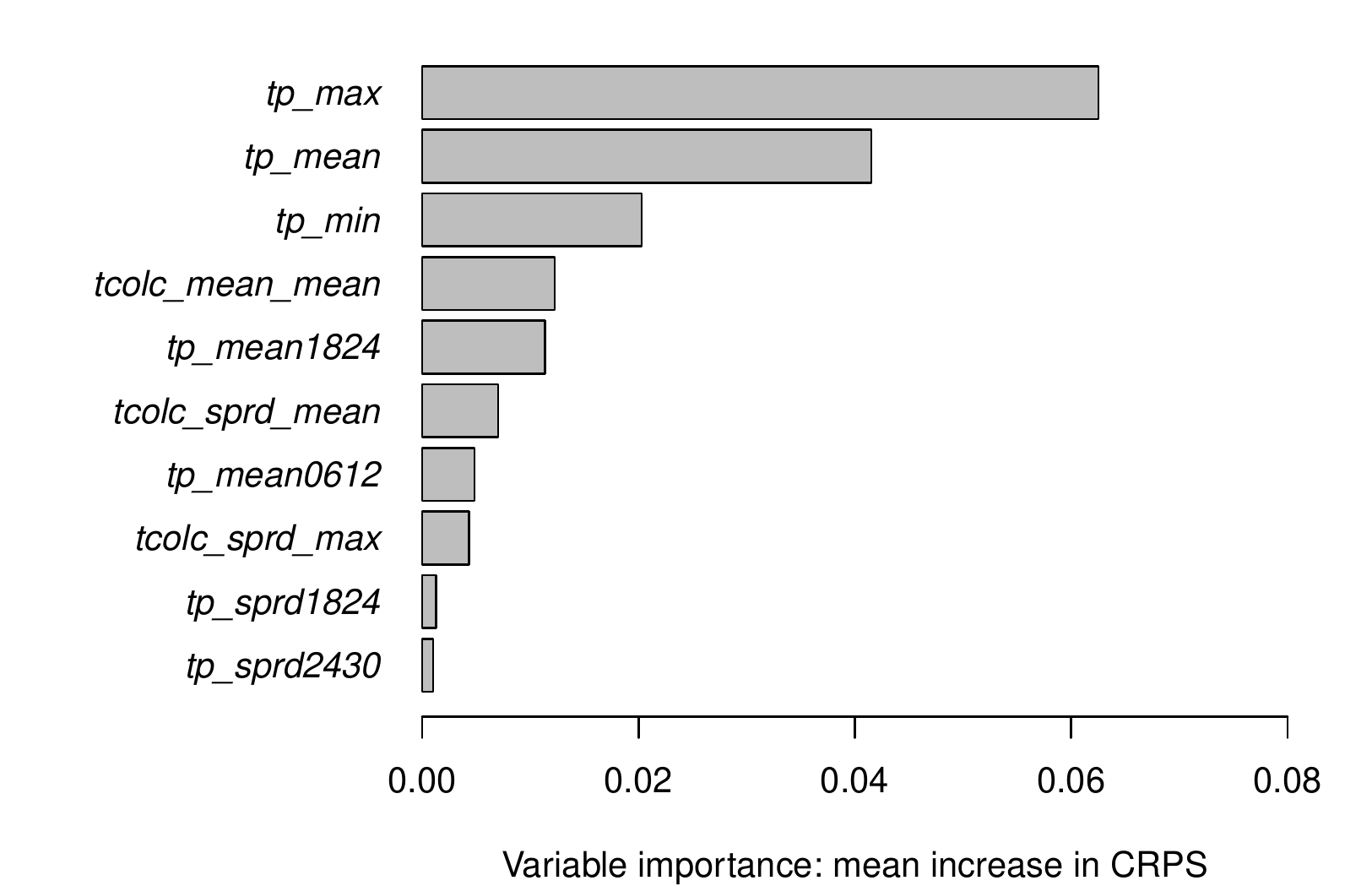}
\caption{\label{fig:axams-varimp} CRPS-based variable importance for the top 10 covariates 
in the distributional forest. Based on data for station Axams, learning period 
1985--2008 and assessed in 2009--2012.}
\end{figure}

To assess the predictive performance, a full cross-validation is carried out
rather than relying on just the one fixed test set for the years 2009--2012.
To do so, a 10 times 7-fold cross-validation is carried out where each repetition
splits the available 28~years into 7~subsets of 4~randomly-selected 
(and thus not necessarily consecutive) years.
The models are learned on 6~folds (= 24~years) and evaluated on the 7-th fold
(= 4~years) using the average CRPS across all observations.
The resulting 10~CRPS skill scores are displayed by boxplots in
Figure~\ref{fig:axams-crps} using EMOS as the reference model (horizontal
line at a CRPSS of 0). Both GAMLSS models and the distributional forest 
perform distinctly better than the EMOS model. While the two GAMLSS lead to an improvement 
of around 4~percent, the distributional forest has a slightly higher improvement
of around 5.5~percent in median.

Finally, it is of interest how this improvement in predictive performance
by the distributional forest is accomplished, i.e., which of the 80~covariates
are selected in the trees of the forest. As the 100~trees of the forest
do not allow to simply assess the variables' role graphically, a common
solution for random forests in general is to consider variable importance
measures. Here, this is defined as the amount of change in CRPS when
the association between one covariate and the response variable is
artificially broken through permutation (and thus also breaking the
association to the remaining covariates).

Figure~\ref{fig:axams-varimp} shows the 10~covariates with the highest permutation
importance (i.e., change in CRPS) for station Axams. As expected the NWP outputs
for total precipitation (\emph{tp}) are particularly important along with total
column-integrated condensate (\emph{tcolc}). Also, both variables occur
in various transformations such as means (either of the full day or certain parts
of the afternoon), spreads, or minima/maxima. Thus, while the covariates
themselves are not surprising, selecting a GAMLSS with a particular combination
of all the transformations would be much more challenging.


\subsection{Application for all stations}
\label{sec:all}

\begin{figure}[t!]
\centering
\includegraphics{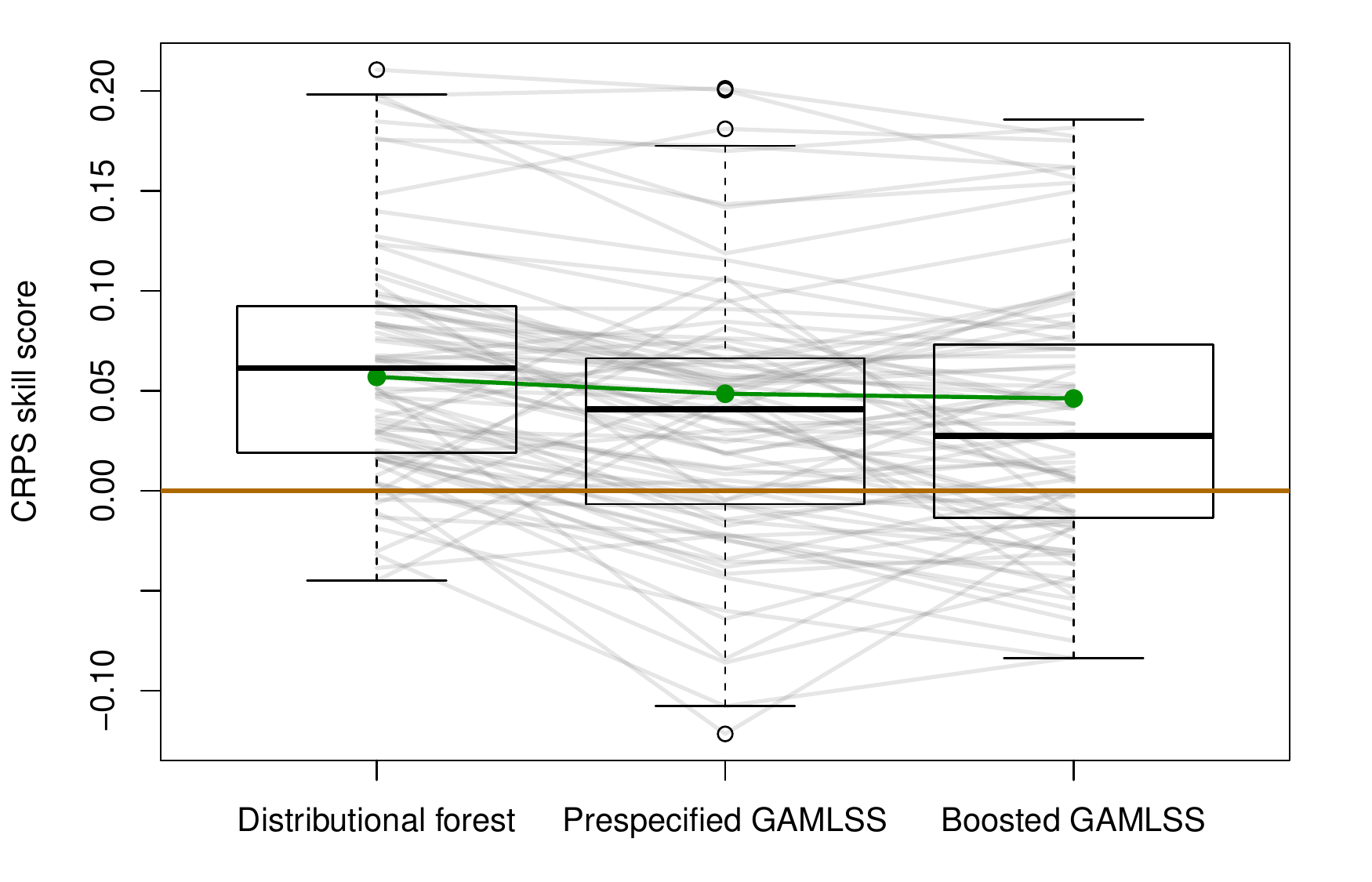}
\caption{\label{fig:all-crps}CRPS skill score for each station (gray lines
with boxplots superimposed). Station Axams is highlighted in green 
and the horizontal orange line pertains to the reference model EMOS. The models are 
learned on 1985--2008 and validated for 2009--2012.}
\end{figure}

\begin{figure}[t!]
\centering
\setkeys{Gin}{width=0.99\textwidth}
\includegraphics{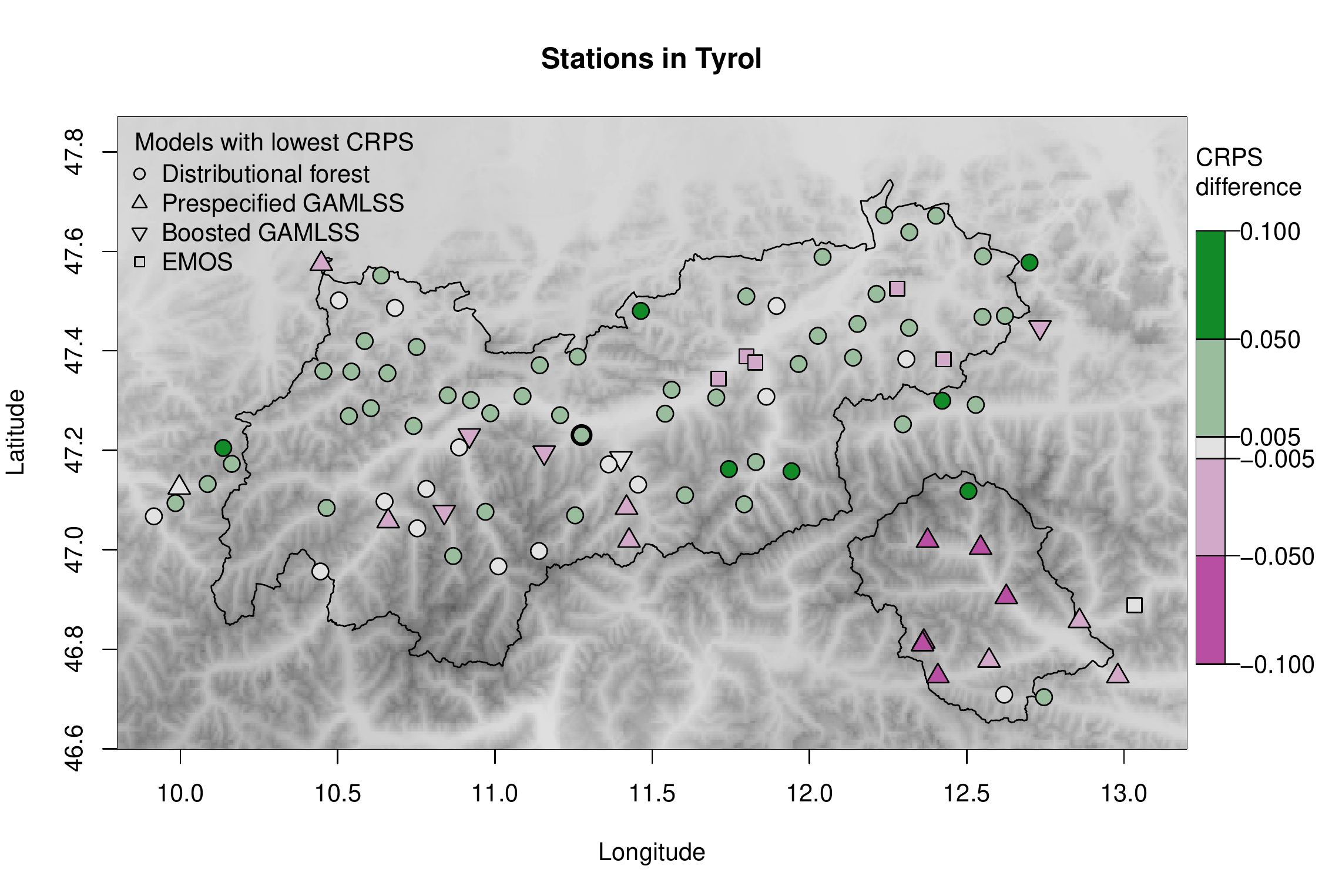}
\caption{\label{fig:all-map}Map of Tyrol coding the best-performing model
for each station (type of symbol) when learned on 1985--2008 and validated for 2009--2012.
The color codes whether the distributional 
forest had higher (green) or lower (red) CRPS compared to the best of the 
other three models. The gray background shows the local topography \citep{Robinson:2014}.
Station Axams is highlighted in bold.}
\end{figure}

After considering only one observational site up to now, a second step
evaluates and compares the competing methods on all 95 available stations. As in the previous
section, all models are learned on the first 24~years and evaluated by the
average CRPS on the last 4~years. More specifically, the CRPS skill score
against the EMOS model is computed for the out-of-sample predictions at each
station and visualized by parallel coordinates plots with boxplots superimposed
in Figure~\ref{fig:all-crps}. Overall, distributional forests have a slightly
higher improvement in CRPSS compared to the two GAMLSS which is best seen
by looking at the boxplots and the green line representing the results
for station Axams. The underlying parallel coordinates additionally
bring out that the prespecified GAMLSS sometimes performs rather differently
(sometimes better, sometimes worse) compared to the two data-driven models.
Values below zero show that, for some stations, EMOS performs better
than the more complex statistical methods.

To assess whether these differences in predictive performance are due to differences
in the topography, Figure~\ref{fig:all-map} shows a brief spatial summary of all
stations. Each station is illustrated by a symbol that conveys which
model performed best in terms of CRPS on the last 4~years of the data.
Additionally, the color of the symbol indicates the CRPS difference between 
distributional forest and the best-performing other model. Green signals that
the distributional forest performs better than the other models whereas red
signals that another model performs better. Overall the distributional forest
performs on par (gray) or better (green) for the majority of stations. Only
for a few stations in the north-east EMOS performs best, and in East Tyrol
the prespecified GAMLSS performs particularly well in the validation period
(2009--2012). Partially, this can be attributed to random variation as
the differences at several stations are mitigated when considering a full
cross-validation rather than a single split into learning and validation period
\citep[see Supplement~B,][and the corresponding discussion in the next section]{Schlosser+Hothorn+Stauffer:2019b}.
Further differences are possibly due to East Tyrol lying in a different
climate zone, south of the main Alpine Ridge. Hence, long-term climatological
characteristics as well as the precipitation patterns in 2009--2012 differ
from North Tyrol, conforming particularly well with the additive effects from
the prespecified GAMLSS.

\section{Discussion}
\label{sec:discussion}

Distributional regression modeling is combined with tree-based modeling
to obtain a novel and flexible method for probabilistic forecasting.
The resulting distributional trees and forests can capture abrupt and
nonlinear effects and interactions in a data-driven way. By basing the
split point and split variable selection on a full likelihood and
corresponding score function, the trees and forests can not only pick
up changes in the location but also the scale or shape of any distributional
family.

Distributional forests are an attractive alternative when prespecifying or
boosting all possible effects and interactions in a GAMLSS model is
challenging.  Distributional forests are rather straightforward to specify
requiring only little prior subject matter knowledge and also work well in the
presence of many potential covariates. The application to precipitation
forecasting in complex terrain illustrates that distributional forests often
perform on par or even better than their GAMLSS counterparts. Hence, they form
a useful addition to the already available toolbox of probabilistic forecasts
for disciplines such as meteorology.

\subsubsection*{Variable selection}

Generally, there are many possibilities how to specify the variables that are
to be included in a distributional regression model. Especially for a low number
of covariates, the GAMLSS approach offers a powerful framework in which
penalized estimation of both smooth main effects and corresponding interaction surfaces
yields models that often balance good predictive performance with high
interpretability (see for example \citealp{Wood+Scheipl+Faraway:2013}; 
\citealp{Goicoa+Adin+Ugarte:2018}; \citealp{Ugarte+Adin+Goicoa:2017}).
However, if the number of covariates is high, including all (or many)
main effects and interactions in a GAMLSS typically becomes challenging both in terms
of interpretability and computational complexity/stability
\citep[see also][]{Hofner+Mayr+Schmid:2016}.

In the precipitation forecasting application, as presented in 
Section~\ref{sec:precipitation}, 80~covariates are considered which corresponds 
to 3160~potential pairwise interactions (and even more higher-order interactions). 
Therefore, only main effects are considered for the boosted GAMLSS while
the prespecified GAMLSS also includes selected interactions chosen based on
meteorological expert knowledge. In contrast, the distributional forest requires no
prespecification as covariates and corresponding interactions are selected automatically.
Thus, distributional forests are an appealing alternative to (boosted) GAMLSS
in weather forecasting tasks as the main concern is typically not so much
interpretability but forecasting skill and (semi-)automatic application on a
larger domain \citep[see also the discussion in][]{Rasp+Lerch:2018}.

\subsubsection*{Distributional specifications for precipitation modeling}

Choosing an adequate distributional family is an important step for
establishing a well-fitting model. A zero-censored Gaussian distribution is
employed in this manuscript as this has been found to be an appropriate choice
for precipitation modeling in earlier literature
\citep[e.g.,][]{Stauffer+Mayr+Messner:2017}. 
To test for robustness against distributional misspecification, 
two alternative distributional specifications have been considered
in Supplement~A \citep{Schlosser+Hothorn+Stauffer:2019a}: Using the
same evaluations as in Section~\ref{sec:all}, all models are
additionally fitted for 15~meteorological stations using a zero-censored
logistic distribution in order to account for heavier tails and a two-part
Gaussian hurdle model combining a binary model for zero vs.~positive
precipitation and a separate Gaussian model, truncated at zero, for the
positive precipitation observations. Both specifications yield qualitatively 
similar results as for the zero-censored Gaussian distribution. For some stations
the two-part hurdle model leads to small improvements, however at the expense
of increased variability across stations (especially for EMOS and the 
boosted GAMLSS). Overall, the results from this manuscript are quite
robust across these distributional specifications, especially for the
distributional forests.

Moreover, one could consider a distribution including an additional parameter
for capturing skewness \citep[as in][]{Scheuerer+Hamill:2015, Baran+Nemoda:2016}. 
However, this would go beyond the mean/variance specification of the NGR
that is widely used in ensemble post-processing. Therefore, this contribution
investigates the effects of using the same distributional family with a novel
strategy for specifying dependence on covariates.

\subsubsection*{More general distributional specifications}

Beyond the task of modeling precipitation it is of interest how well distributional
forests perform in combination with other more general distributional specifications.
It has been shown previously in the literature that using a score- or gradient-based
selection of splitting variables outperforms a mean-based selection with subsequent
flexible distributional modeling: For example, both \citet[][Figure~2]{Athey+Tibshirani+Wager:2019}
and \citet[][Figure~1]{Hothorn+Zeileis:2017} demonstrated (independently) that their
respective score-based random forest algorithms outperform the mean-based quantile
regression forests of \cite{Meinshausen:2006} in a setup where only the variance of a normal
response variable changes across the considered covariates. However, if all distribution
parameters are closely correlated with the distribution mean the forests with different
splitting strategies all perform similarly, provided a sufficiently flexible distribution
is employed for the final predictions \citep[see][Section~7]{Hothorn+Zeileis:2017}.

Similarly, the score-based distributional forests introduced in this manuscript
proved to be quite robust to the different distributional specifications considered.
While all specifications focus on capturing mean-variance effects note that these
parameters are never fully orthogonal but can actually become quite closely correlated
due to the censoring (or truncation and/or zero-inflation considered in
Supplement~A \citep{Schlosser+Hothorn+Stauffer:2019a}).

However, exploring extensions to more flexible parametric distributions
\citep[e.g., such as the Dagum distribution considered by][in GAMLSS-type models]{Klein+Kneib+Lang:2015}
as well as transformation model specifications \citep[e.g., as in][]{Hothorn+Zeileis:2017}
are of interest for future research.

\subsubsection*{Axams vs.~other meteorological stations}

Axams was chosen as the meteorological station for the more extensive evaluations
in Section~\ref{sec:axams} as it yields fairly typical results
and is geographically in the center of the study area and closest to
Innsbruck, the capital of Tyrol and the work place of three of the authors.
To show that qualitatively similar results are obtained for other
meteorological stations, Supplement~B \citep{Schlosser+Hothorn+Stauffer:2019b}
carries out the same evaluation for 14~further stations. These
cover a wide range of geographical locations/altitudes and a mix of
different best-performing models in the single-split setting reported in
Section~\ref{sec:all}.

The supplement shows that some of the differences in forecast skill from
Figure~\ref{fig:all-map} even out in the cross-validation with distributional
forests typically performing at least as well as the best of the other models at most
stations. In particular, this also includes three stations in East Tyrol
where the prespecified GAMLSS performs best in the single-split setting
(learning based on 1985--2008 and validation for 2009--2012).

\subsubsection*{Tuning parameters}

Selecting tuning parameters for flexible regression models is important
not only in terms of predictive accuracy but also computational complexity. 
For the application in Section~\ref{sec:precipitation} tuning parameters
are selected based on advice from the literature as well as our own experiences.
As \cite{Hastie+Tibshirani+Friedman:2001} and \cite{Breiman:2001} recommend to 
build full-grown trees, early stopping upon non-significance is disabled
(\texttt{alpha}~$= 1$) and low values are used for \texttt{minsplit}~($ = 50$)
and \texttt{minbucket}~($ = 20$), while assuring that \texttt{minsplit}
is sufficiently large for reasonably obtaining MLEs of all parameters in each
segment of the tree.

Applying the Law of Large Numbers it can be shown that random forests do not overfit 
as the number of trees increases
\citep{Breiman:2001,Hastie+Tibshirani+Friedman:2001,Biau+Scornet:2016}.
Therefore, in principle, forests can be built with a very large number of trees
(\texttt{ntree}) as this cannot deteriorate the predictions. However, ``[\dots] the
computational cost for inducing a forest increases linearly with the number of trees,
so a good choice results from a trade-off between computational complexity and accuracy''
\citep[][p.~205]{Biau+Scornet:2016}.
Following this advice, we decided to build forests consisting of 100 trees.

\subsubsection*{Computational difficulties}
As stated by \cite{Hofner+Mayr+Schmid:2016} the AIC-based variable selection 
methods implemented in the \textsf{R} package \textbf{gamlss} ``[...] can be 
unstable, especially when it comes to selecting possibly different sets of variables for 
multiple distribution parameters.'' We have noticed computational problems when
applying \textbf{gamlss} in certain settings within the cross-validation
framework as it did not succeed in fitting the model. In these cases the prespecified
GAMLSS was not taken into consideration in the comparison of all applied models.

\section*{Computational details}

The proposed methods are in the \textsf{R} package \textbf{disttree}
(version~0.1.0) based on the \textbf{partykit}
package (version~1.2.3), both available 
on \textsf{R}-Forge at
(\url{https://R-Forge.R-project.org/projects/partykit/}). 
The function \code{distforest} learns the distributional forests proposed
in this manuscript by combining the general \code{cforest} function from
\textbf{partykit} with the function \code{distfit} for fitting distributional
models by maximum likelihood. Analogously, \code{disttree} can learn a single
distributional tree by combining \code{ctree} with \code{distfit}.
All functions can either be used with GAMLSS family objects from the \textsf{R} package \textbf{gamlss.dist} 
(\citealp{Stasinopoulos+Rigby:2007}, version~5.0.6)
or with custom lists containing all required information about the distribution family.

In addition to \textbf{disttree}, Section~\ref{sec:precipitation} employs
\textsf{R} package \textbf{crch} 
\citep[version~1.0.1]{Messner+Mayr+Zeileis:2016} for the EMOS 
models, \textbf{gamlss}
\citep[version~5.1.0]{Stasinopoulos+Rigby:2007} for the 
prespecified GAMLSS, and \textbf{gamboostLSS}
\citep[version~2.0.1]{Hofner+Mayr+Schmid:2016} for the boosted GAMLSS.

The fitted distributional forest for July 24 and
observation station Axams (including Figure~\ref{fig:axams-dist})
is reproducible using \code{demo("RainAxams", package = "disttree")}.
This also includes fitting the other zero-censored
Gaussian models considered in this paper and generating the corresponding 
QQ~plots (Figure~\ref{fig:axams-qq}) and PIT histograms 
\citep[Supplement~B]{Schlosser+Hothorn+Stauffer:2019b}.
Full replication of all results can be obtained with
\code{demo("RainTyrol", package = "disttree")}
requiring the companion \textsf{R} package \textbf{RainTyrol} 
(version~0.1.0),
also available within the \textsf{R}-Forge~project.
The results presented in Supplement~A \citep{Schlosser+Hothorn+Stauffer:2019a} and 
Supplement~B \citep{Schlosser+Hothorn+Stauffer:2019b} can be reproduced using
\code{demo("RainDistributions", package = "disttree")} and 
\code{demo("RainStationwise", package = "disttree")}, respectively.

\section*{Appendix}
\begin{appendix}

\section{Tree algorithm}
\label{app:tree}

In the following, the tree algorithm applied in the empirical case study 
discussed in this paper is explained. For notational simplicity, 
the testing and splitting procedure is described for the root node, i.e.,
the entire learning sample with observations $\{y_i\}_{i = 1,\ldots,n}$, $n \in \mathbb{N}$.
In each child node the corresponding subsample depends on 
the foregoing split(s).

After fitting a distributional model $\mathcal{D}(Y, \bm{\theta})$ to 
the learning sample with observations $\{y_i\}_{i = 1,\ldots,n}$ as explained in 
Section~\ref{sec:distfit} the resulting estimated parameter 
$\bm{\hat{\theta}} = 
(\hat{\theta}_1, \ldots, \hat{\theta}_k)$, $k \in \mathbb{N}$ 
can be plugged in the score function $s(\bm{\theta}, Y)$.
In that way a goodness-of-fit measurement is obtained for each
parameter $\theta_j$ and each observation $y_i$.
To use this information, statistical tests are employed to detect
dependencies between the score values
\begin{equation}
s(\bm{\hat{\theta}}, y) = 
\begin{pmatrix} 
s(\bm{\hat{\theta}}, y_1)_1 & s(\bm{\hat{\theta}}, y_1)_2 & \ldots & s(\bm{\hat{\theta}}, y_1)_k\\
\vdots & \vdots & \ddots & \vdots \\
s(\bm{\hat{\theta}}, y_n)_1 & s(\bm{\hat{\theta}}, y_n)_2 & \ldots & s(\bm{\hat{\theta}}, y_n)_k
\end{pmatrix}
\end{equation}
and each variable $Z_l \in \{Z_1, \ldots, Z_m\}$.  
More formally, the following hypotheses are assessed with permutation tests:
\begin{align}
H_0^l:  s(\bm{\hat{\theta}}, Y) \qquad \bot \qquad Z_l.
\end{align}
The permutation tests are based on the multivariate linear statistic
\begin{equation}
T_l = vec\left(\sum_{i=1}^n v_l(Z_{li}) \cdot s(\bm{\hat{\theta}}, Y_i)\right)
\end{equation}
where $s(\bm{\hat{\theta}}, Y_i) \in \mathbb{R}^{1\times k}$ and the 
type of the transformation function $v_l$ depends on the type of the split 
variable $Z_l$. If $Z_l$ is numeric then $v_l$ is simply the identity 
function $v_l(Z_{li}) = Z_{li}$ and therefore $T_l \in \mathbb{R}^k$ as 
the ``vec'' operator converts the $1 \times k$ matrix into a $k$ column vector. 
If $Z_l$ is a categorical variable with $H$ categories then 
$v_l(Z_{li}) = (\I(Z_{li} = 1), \ldots, \I(Z_{li} = H))$ 
such that $v_l$ is a $H$-dimensional unit vector where the element corresponding to 
the value of $Z_{li}$ is $1$. In this case the statistic $T_l \in \mathbb{R}^{H \cdot k}$
as the ``vec'' operator converts the $H \times k$ matrix into a $H \cdot k$ column vector 
by column-wise combination. Observations with missing values are excluded from the sums.

With the conditional expectation $\mu_l$ and the covariance 
$\Sigma_l$ of $T_l$ as derived by 
\cite{Strasser+Weber:1999} the test statistic can be standardized. 
The observed multivariate linear statistic $t_l$ which is either a
$k$- or $k \cdot H$-dimensional vector, depending on the scale of $Z_l$,
is mapped onto the real line by a univariate test statistic $c$.
In the application of this paper a quadratic form is chosen, such that
\begin{equation}
c_{\text{quad}}(t_l,\mu_l,\Sigma_l) = (t_l-\mu_l)\Sigma_l^+(t_l-\mu_l)^{\top}
\end{equation}
where $\Sigma_l^+$ is the Moore-Penrose inverse of $\Sigma_l$. Alternatively,
the maximum of the absolute values of the standardized linear statistic
can be considered ($c_{\text{max}}$).

\cite{Strasser+Weber:1999} showed that the asymptotic conditional 
distribution of the linear statistic $t_l$ is a multivariate normal with parameters
$\mu$ and $\Sigma$. Hence, the asymptotic conditional distribution of
$c(t_{l},\mu_{l},\Sigma_{l})$ 
is either normal
(for $c_{\text{max}}$) 
or 
$\chi^2$ (for $c_{\text{quad}}$).

The smaller the $p$-value corresponding to the standardized test 
statistic $c(t_{l},\mu_{l},\Sigma_{l})$ is the stronger the discrepancy 
from the assumption of independence between the scores and the split 
variable $Z_l$.
After Bonferroni-adjusting the $p$-values it has to be assessed whether
any of the resulting $p$-values is beneath the selected significance level. 
If so, the partitioning variable $Z_{l^\ast}$ with the lowest $p$-value 
is chosen as splitting variable. Otherwise no further split is made 
in this node as the stopping criterion of no $p$-values being below the 
significance level is fulfilled. This type of early-stopping in building
a tree is sometimes also referred to as ``pre-pruning''. For random forests
pre-pruning is often switched off by setting the significance level to $1$.

The breakpoint that leads to the highest discrepancy between score functions 
in the two resulting subgroups is selected as split point. 
This is measured by the linear statistic
\begin{equation}
T_{l^{\ast}}^{qr} = \sum_{i \in \mathcal{B}_{qr}} s(\bm{\hat{\theta}}, Y_i)
\end{equation}
for $q \in \{1,2\}$ where $\mathcal{B}_{1r}$ and $\mathcal{B}_{2r}$ are the 
two new subgroups, without any particular ordering, that are defined by 
splitting in split point $r$ of variable $Z_{l^{\ast}}$.
The split point is then chosen as follows:
\begin{equation}
r^{\ast} = \argmin_{r} (\min_{q=1,2}(c(t_{l^{\ast}}^{qr},\mu_{l^{\ast}}^{qr},\Sigma_{l^{\ast}}^{qr})).
\end{equation}
One repeats the testing and splitting procedure in each of the resulting 
subgroups until some stopping criterion is reached. This criterion 
can for example be a minimal number of observations in a node or a 
minimal $p$-value for the statistical tests. In that way pre-pruning
is applied in order to find right-sized trees and hence 
avoid overfitting.

This permutation-test-based tree algorithm is presented in 
\cite{Hothorn+Hornik+Zeileis:2006} as the CTree algorithm. 
A different framework to build a likelihood-based tree is provided by the MOB algorithm 
which is based on M-fluctuation tests (\citealp{Zeileis+Hothorn+Hornik:2008}).




\end{appendix}

\section*{Acknowledgments}
An extended research stay of Torsten Hothorn in Innsbruck (August
2017 to January 2018) was financially supported by the Swiss National
Science Foundation, grant number SNF IZSEZ0 177091.

\section*{Supplementary material}
\subsection*{Supplement~A: Different Response Distributions}
  To assess the goodness of fit of the Gaussian distribution, left-censored at zero,
  this supplement employs the same evaluations as in the main manuscript but based on
  two other distributional assumptions: A logistic distribution, left-censored at zero,
  is employed to potentially better capture heavy tails -- and a two-part hurdle model
  combining a binary model for zero vs.\ positive precipitation and a Gaussian model,
  truncated at zero, for the positive precipitation observations. 

\subsection*{Supplement~B: Stationwise Evaluation}
  To show that Axams is a fairly typical
  station and similar insights can be obtained for other stations as well,
  this supplement presents the same analysis as in Section~\ref{sec:axams} 
  of the main manuscript for 14~further meteorological stations.

\bibliography{ref.bib}

\end{document}